**Hedging American Put Options with Deep Reinforcement Learning**

**Reilly Pickard [1 *], Finn Wredenhagen[2], Julio DeJesus[2], Mario Schlener[2], Yuri Lawryshyn [3]**

**April 2024**

1 Department of Mechanical and Industrial Engineering, University of Toronto, Toronto, ON M5S 3G8, Canada
2 Ernst & Young LLP, Toronto, ON, M5H 0B3, Canada
3 Department of Chemical Engineering and Applied Chemistry, University of Toronto, Toronto, ON M5S 3E5, Canada
* Correspondence: reilly.pickard@mail.utoronto.ca

**Hedging American Put Options with Deep Reinforcement Learning**

**April 2024**

**ABSTRACT:** This article leverages deep reinforcement learning (DRL) to hedge American put options, utilizing the deep deterministic policy gradient (DDPG) method. The agents are first trained and tested with Geometric Brownian Motion (GBM) asset paths and demonstrate superior performance over traditional strategies like the Black-Scholes (BS) Delta, particularly in the presence of transaction costs. To assess the real-world applicability of DRL hedging, a second round of experiments uses a market calibrated stochastic volatility model to train DRL agents. Specifically, 80 put options across 8 symbols are collected, stochastic volatility model coefficients are calibrated for each symbol, and a DRL agent is trained for each of the 80 options by simulating paths of the respective calibrated model. Not only do DRL agents outperform the BS Delta method when testing is conducted using the same calibrated stochastic volatility model data from training, but DRL agents achieves better results when hedging the true asset path that occurred between the option sale date and the maturity. As such, not only does this study present the first DRL agents tailored for American put option hedging, but results on both simulated and empirical market testing data also suggest the optimality of DRL agents over the BS Delta method in real-world scenarios. Finally, note that this study employs a model-agnostic Chebyshev interpolation method to provide DRL agents with option prices at each time step when a stochastic volatility model is used, thereby providing a general framework for an easy extension to more complex underlying asset processes.

# Introduction

## BACKGROUND

Following the sale of a financial option, traders seek to mitigate the associated risk through hedging strategies. For example, a trader that has just sold a European put option may hedge against adverse price drops by selling shares in the underlying asset. Traditionally, option hedgers attempt to maintain a Delta neutral portfolio, where Delta is the first partial derivative of the option price with respect to the underlying asset (Hull 2012). Therefore, the Delta neutral hedge for the European put option requires the sale of Delta shares of the underlying. For European options, analytic solutions for the option price and Delta exist via the Black and Scholes (BS) option pricing model (Black and Scholes 1973). While the BS model shows that the underlying risk of an option position is eliminated by a continuously rebalanced Delta-neutral portfolio, financial markets operate discretely in practice. Further, the BS model assumes constant volatility and no trading costs, which is not reflective of reality. As such, dynamic option hedging under market frictions is a progressive decision-making progress under uncertainty. One field that has garnered significant attention in dynamic decision-making procedures is reinforcement learning (RL), a subfield of artificial intelligence (AI). RL problems are aided in complex environments through neural network (NN) function approximation. The field concerning the combination of NN's and RL is called deep RL (DRL), and DRL has been used to achieve super-human level performance in video games (Mnih et al. 2013), board games (Silver et al. 2016), and robot control (Lillicrap et al. 2015).

As the dynamic hedging problem requires decision-making under uncertainty, several recent studies have used DRL to effectively hedge option positions. A review of 17 studies that use DRL for dynamic stock option hedging is given by Pickard and Lawryshyn (2023), who detail that while many studies show that DRL outperforms a Delta-neutral strategy when hedging European options under transaction costs and stochastic volatility, no current studies consider the hedging of American options. For American put options specifically, there is no analytical pricing or hedging formula available due to the potential of early exercise, and numerical methods for option pricing and hedging are required (Hull 2012). To address the literature gap pertaining to DRL agents that consider the potential for early option exercise, this article details the design of DRL agents that are trained to hedge American put options under transaction costs. DRL agents in this study are designed using the deep deterministic policy gradient (DDPG) method. In addition to training an American put DRL hedger when the underlying asset price follows a geometric Brownian motion (GBM), stochastic volatility is considered by calibrating stochastic volatility models using empirical option data on several stock symbols. Once these DRL agents are tested on simulated paths generated by the calibrated model, the hedging performance of each DRL agent is evaluated on the empirical asset price path for the respective symbol between the sale and maturity dates.

Finally, note that DRL agent reward function requires the option price at each time step. While an interpolation of a binomial American option tree is used in GBM cases, this study employs the use of a Chebyshev interpolation method first proposed by Glau, Mahlstedt, and Potz (2018) for the determination of the option price in stochastic volatility experiments. This Chebyshev method is model-agnostic, and this work thereby provides a framework that extends seamlessly to more intricate processes. Moreover, the Chebyshev method allows the American option price to be computed more efficiently in stochastic volatility settings, as the requirement to average the payoff of several thousand Monte Carlo (MC) simulations from the current price level to expiry or exercise is eliminated. This Chebyshev pricing method is described in detail in the methodology section of this work.

The rest of this section is dedicated to the introduction of DRL and a detailed account of similar work in the DRL hedging space. This article will then detail the methodology used to train DRL agents, before presenting and discussing the results of all numerical experiments.

## REINFORCEMENT LEARNING

RL, which is given a thorough explanation by Sutton and Barto (2018), is summarized in this section. RL is a continuous feedback loop between an agent and its environment. At each time step, the RL agent, in a current state $s$, takes an action $a$, and receives a reward $r$. The objective is to maximize future expected rewards, $G_t$, which, at time step $t$ for an episodic task of $T$ discrete steps, are computed as

$$G_t = r_{t+1} + \gamma r_{t+2} + \gamma^2 r_{t+3} \dots + \gamma^{T-1} r_T, \tag{1}$$

noting that $\gamma$ is the discount factor to avoid infinite returns. For any state $s,$ the corresponding action is given by the agent's policy, $\pi(s) \rightarrow a$. Policies are evaluated by computing value functions such as the $Q$-function, which maps state-action pairs to their expected returns under the current policy, i.e.,

$$Q^{\pi} = Q^{\pi}(s,a) = \mathbb{E}[G|a,s] \in \mathbb{R}, \forall\, s \in \mathcal{S}, \forall\, a\, \mathcal{A}. \tag{2}$$

$\mathcal{S}$ and $\mathcal{A}$ are the state and action-spaces, respectively. To maximize the expected return from all states in $\mathcal{S}$, the RL agent aims to learn the $Q$-function stemming from the optimal policy, $Q^{\pi^*}(s,a)$. The $Q$-function may be updated iteratively in an on- or off-line manner, with both on- and off-policy methods. An example of an on-line, on-policy method is temporal-difference (TD) learning, which updates the $Q$-value after each action. For example, the on-policy TD SARSA $(s, a, r', s', a')$ algorithm updates the $Q$-value for a given state-action pair as

$$Q(s,a) \leftarrow Q(s,a) + \alpha(r' + \gamma Q(s',a') - Q(s,a)), \tag{3}$$

where $r'$ is the reward received after taking action $a$ in state $s$, $(s', a')$ is the state-action pair at the subsequent time step, and $\alpha$ is the step-size for the update. As the $Q$-function is updated, the policy is improved. In traditional value-based methods, policy improvement requires making the current policy greedy with respect to the $Q$-function, which, for each state, requires an argmax operation over all actions that maximize a given $Q$-value. On-policy agents learn about the environment while following this improved policy. As such, on-policy agents may under-explore potentially optimal actions, even if there is some built-in probability that the next action is random as opposed to greedy. Conversely, off-policy methods involve using a sub-optimal policy to generate experiences for updating $Q^{\pi^*}(s,a)$, allowing for more exploration. A breakthrough result for off-policy RL stems from Watkins (1989), who introduced $Q$-learning, which updates the current Q-value as

$$Q(s,a) = Q(s,a) + \alpha\big(r' + \gamma max_a Q(s',a') - Q(s,a)\big). \tag{4}$$

SARSA and $Q$-learning are tabular methods, which require storing the $Q$-value for each state-action pair and policy for each state in a lookup table. As a tabular $Q$-function requires the storage of $\mathcal{S} \times \mathcal{A}$ values, high-dimensional problems become intractable. However, the $Q$-function may be approximated with function approximators such as NNs, recalling that this combination of NNs and RL is called DRL. DRL was popularized by the work of Mnih et al. (2013), who approximate the $Q$-function with a NN, specifically denoting each $Q$-value by $Q(s, a;\ \theta)$, where $\theta$ is a vector of the NN parameters that signifies the weights and biases connecting the nodes within the neural network. Mnih et al. (2013) show that this method of deep Q-networks (DQN) can be used to train an agent to master multiple Atari games.

The Q-network parameters stored in $\theta$ are optimized by minimizing a loss function between the Q-network output and the target value. The loss function for iteration $i$ is given by

$$L_i(\theta_i) = \mathbb{E}_{s,a}\left[\left(y_i - Q(s, a;\ \theta_i)\right)^2\right], \tag{5}$$

where $y_i = \mathbb{E}_{s,a}\ [r' + \gamma max Q(s', a';\ \theta_{i-1})]$ is the target $Q$-value for iteration $i$ (Mnih et al. 2013). Based on this loss value, $\theta$ is updated in the direction of the negative gradient of the loss function with respect to $\theta$, which is the method of stochastic gradient descent (SGD) (Mnih et al. 2013). SGD updates the weights as

$$\theta_i = \theta_{i-1} - \eta \nabla_{\theta_i} L_i(\theta_i), \tag{6}$$

where $\eta$ is the learning rate for the neural network, and $\nabla_{\theta_i} L_i(\theta_i)$ is the gradient of the loss with respect to the network weights. In addition to $Q$-network approximation, another component of DQN is the use of the experience replay buffer (Mnih et al. 2013). Experience replay, originally proposed for RL by Lin (1992), stores encountered transitions (state, action, and reward) in a memory buffer. Mnih et al. (2013) use the replay buffer to uniformly sample the target $Q$-value for iteration $i$ ($y_i$).

To update the Q-function in value methods such as DQN, a max operation on the $Q$-value over all next possible actions is required. Further, as discussed above, improving the policy in value methods requires an argmax operation over the entire action-space. When the action-space is continuous, the valuation of each potential next action becomes intractable. In financial option hedging, discretizing the action-space restricts the available hedging decisions. While hedging does require the acquisition of an integer number of shares, a continuous action-space for the optimal hedge provides more accuracy, as the hedge is not limited to a discrete set of choices. As such, continuous action-spaces are much more prevalent in the DRL hedging literature (Pickard and Lawryshyn 2023)

To handle problems with continuous action-spaces, Lillicrap et al. (2015) introduce a policy-based algorithm called DDPG, that utilizes an actor-critic architecture. For each action, a $Q$-value is computed by the critic network, which is trained by SGD (Lillicrap et al. 2015). While this SGD operation on the critic network is akin to DQN, the key change is in the computation of the target $y_i$, which is used in the calculation of the loss, $L_i(\theta_i)$. In DDPG, the target is computed as $y_i = \mathbb{E}_{s,a}\ [r' + \gamma Q(s', \pi_i(s');\ \theta_{i-1}\ )]$ (Lillicrap et al. 2015), noting that the current policy estimate from the actor network, $\pi_i(s')$ is used, rather than a max operation over $a'$ as in DQN. Eliminating the requirement to search all action candidates to compute the optimal target enables the use of a continuous action-space. Further, an optimization of the actor network eliminates the need to perform an argmax operation over all next actions to make the best choice (Sutton and Barto 2018). Once the $Q$-value output from the critic is computed, the $Q$-value is passed back to the actor network, and the actor network is optimized by performing gradient ascent on the current $Q$-value with respect to the actor network weights (Lillicrap et al. 2015).

## SIMILAR WORK

Pioneering works in the field of RL hedging are given by Halperin (2017) and Kolm and Ritter (2019). Halperin (2017) approximates a parametrized optimal $Q$-function with fitted $Q$-learning, and Kolm and Ritter (2019) use a function-approximation SARSA method by fitting a non-linear function of the form $y = Q(X)$ to approximate the $Q$-function, where $X$ is the current state-action pair $(s, a)$. In terms of those who use DRL to solve the option hedging problem, Du et al. (2020) and Giurca and Borovkova (2021) use DQN to approximate the optimal $Q$-function. Studies employing the policy method of DDPG include Cao et al. (2021), Assa, Kenyon, and Zhang (2021), Xu and Dai (2022), and Fathi and Hientzsch (2023). Cao et al. (2023) extend DDPG in their DRL hedging study, as instead of estimating the $Q$-value, the critic network uses quantile regression (QR) to estimate the distribution of discrete fixed-point rewards.

In the DRL hedging literature, most studies include the current asset price, time-to-maturity, and current holding in the state-space (Pickard and Lawryshyn 2023). Vittori et al. (2020), Giurca and Borovkova (2021), Xu and Dai (2022), Fathi and Hientzch (2023), and Zheng, He, and Yang (2023) include the BS Delta in the state-space for their study, while several papers explicitly explain that the BS Delta can be deduced from the option and stock prices and its inclusion only serves to unnecessarily augment the state (Cao et al. (2021), Kolm and Ritter (2019), Du et al. (2020)). As for the reward function, the mean-variance reward formulation is commonly used to provide feedback to the agent at each time step. A generalized version of the mean-variance reward for a single time-step $t$ is given as

$$r_t = \delta w_t - \xi \delta {w_t}^2, \tag{7}$$

where $\xi$ is a measure of risk aversion, and $\delta w_t$ is the incremental wealth of the hedging portfolio at time $t$. Note the wealth increment is given by $\delta w_t = [C_t - C_{t-1}] + n_t[S_t - S_{t-1}] - c_t$, where $C_t$, is the option value, $n_t$ is the position in the underlying, and $c_t$ is a transaction cost. The mean-variance reward formulation is used in Halperin (2017, 2019), Kolm and Ritter (2019), Du et al. (2020), Giurca and Borokova (2021), Xu and Dai (2022), Canelli et al. (2023), and Zheng, He, and Yang (2023). To train DRL agents to monitor transaction costs when hedging, several papers include a transaction-cost penalty to the reward (Pickard and Lawryshyn 2023).

Given a defined state-space, action-space, and reward, a DRL agent needs experience in the form of data to learn optimal policies and value functions. In the dynamic hedging environment, the data may be simulated or empirical, i.e., market data. In a simulated case, the underlying asset follows a stochastic process such as GBM or a stochastic volatility model such as the SABR (Stochastic Alpha, Beta, Rho) (Hagan et al. 2002) or Heston (1993) models. Monte Carlo simulations of the GBM process are employed for data generation in most studies in the RL hedging literature (Pickard and Lawryshyn 2023). To train DRL agents for an environment with stochastic volatility, Cao et al. (2021) use a SABR model, and the Heston model is used to generate experimental data in Mikkilä and Kanniainen (2023), Murray et al. (2022), Xiao, Yao, and Zhou (2021), and Giurca and Borovkova (2021). It is noted that no papers in the literature calibrate the stochastic volatility model parameters to market data.

Consistent results emerge across the DRL hedging literature when training and testing on simulated data in that DRL agents trained with a transaction cost penalty outperform the traditional BS Delta hedging strategy in frictional markets (Pickard and Lawryshyn 2023). While Giurca and Borovkova (2021), Xiao, Yao, and Zhou (2021), Zheng, He, and Yang (2023), and Mikkilä and Kanniainen (2023) train their agents on simulated data, each of these studies also performs a sim-to-real test in which they test the simulation trained DRL agent on empirical data. Notably, Xiao, Yao, and Zhou (2021), Mikkilä and Kanniainen (2023), and Zheng, He, and Yang (2023) achieve desirable results in the sim-to-real test, while Giurca and Borovkova (2021) do not. In addition to using empirical data for a sim-to-real test, Mikkilä & Kanniainen (2023) train their DRL agent using the empirical dataset. Others who train a DRL agent with empirical data are Pham, Luu, and Tran (2021) and Xu and Dai (2022). Pham, Luu, and Tran (2021) show that their DRL agent has a profit higher than the market return when testing with empirical prices. Of further note, Xu and Dai (2022) train and test their DRL agent with empirical American option data, which is not seen elsewhere in the DRL hedging literature. However, Xu and Dai (2022) do not change their DRL approach in any manner when considering American versus European options. There have been no encountered studies that train DRL agents specifically to hedge American options by incorporating exercise boundaries into the training and testing process. This observation that there is an absence of literature on DRL for hedging American options is supported by two literature reviews that analyze the use of DRL for dynamic option hedging (Pickard and Lawryshyn 2023, Liu 2023).

## Methodology

Successfully implementing a DRL agent to achieve the American put option hedging task requires three main steps: the design of the DRL agent, the setup of the training procedure, and the setup of a testing procedure. The design of the DRL agent features the construction of the neural network, the choice of hyperparameters, the formulation of state- and action-spaces, and a reward. The training procedure involves the data generation processes required to provide the agent with adequate state and reward information. Finally, forming testing scenarios requires the acquisition of data, and the development of a benchmark comparator for the DRL agent, such as the Delta hedging method. This section will first detail the DRL agent design before detailing the training and testing procedures for all experiments.

### DRL AGENT DESIGN

As policy based DRL methods allow for continuous action-spaces, the DRL method employed in this study is DDPG. In this study, the actor and critic networks are both fully connected NNs with two hidden layers consisting of 64 nodes each. In both the actor and critic networks, the hidden layers use the rectified linear unit as the non-linear activation function. The actor network uses a sigmoidal output function to map the actions to the range [0, 1], while the critic network uses a linear output. Note that the actor output is multiplied by $-1$, as this study hedges American put options, which require shares to be shorted.

An important hyperparameter for NN training is the learning rate. Careful selection must be made to balance efficiency and stability: setting the learning rate too low may lead to overfitting, and setting the learning too high may lead to instability and divergence in the training process (Smith 2018). In this study, satisfactory results stemmed from setting the actor and critic learning rates to $5 \times 10^{-6}$ and $5 \times 10^{-4}$, respectively.

The state-space for the DRL agent in this study includes the current asset price, the time-to-maturity, and the current holding (previous action). This is aligned with the state-spaces used in Kolm and Ritter (2019), Cao et al. (2021), Xiao, Yao, and Zhou (2021), and Assa, Kenyon, and Zhang (2021). This study does not include the BS Delta in the state-space, agreeing with Cao et al. (2021), Kolm and Ritter (2019), and Du et al. (2020) in that the addition of the BS Delta is an unnecessary augmentation of the state. Moreover, the inclusion of the BS Delta in the state may hinder the effectiveness of an American DRL hedger, as the BS model is derived using European options.

The reward formulation used in this study is given by

$$R_t = -|A_t(S_t - S_{t-1}) - (C_t - C_{t-1})| - \kappa(A_t - A_{t-1})^2 S_t. \tag{8}$$

$A_t, S_t$, and $C_t$ represent the hedging action, asset price, and option price at time $t$, while $\kappa$ is the quadratic transaction cost penalty multiplier. Taking the absolute negative difference between the option value change and asset holding change at each time step is reflective of a hedging goal in that in the absence of transaction costs, the optimal hedge should yield a reward of zero. The transaction cost penalty in the reward function differs from the literature. For example, Cao et al. (2021) use a linear transaction cost function, modelled as $\lambda(A_t - A_{t-1})S_t$, where $\lambda$ is the transaction cost rate. However, it was found in this study that the quadratic training penalty led to better results. When using the linear transaction cost penalty, setting $\lambda$ too low led to a non-cost-conscious agent, and making $\lambda$ too high yielded an agent that underhedged at early time steps. While under hedging at early time steps may be satisfactory when hedging European options, American options pose the threat of early payouts, and under hedging leaves the hedger exposed if the option is suddenly exercised. As such, best results were achieved when using a quadratic transaction cost penalty, as large action changes that would yield the highest transaction costs are punished more heavily than the small action changes required to effectively hedge American options. For all DRL agents trained in this study, $\kappa$ is set to $0.005$. While training is performed with a quadratic transaction cost penalty, the transaction costs used for testing are modelled as a linear function of the asset price with the function $\lambda(A_t - A_{t-1})S_t$, which is more reflective of reality and consistent with Cao et al. (2021). Note that all DRL agents in this study are trained over 5000 episodes, each consisting of 25 timesteps, which was found to be a satisfactory amount of data.

## TRAINING PROCEDURES

Given the design of the state, action, and reward, training a DRL agent to hedge American put options requires the following:

1. The generation of asset price data, used in the DRL agent state and reward.
2. The American put option price, used in the DRL agent reward.

It is noted that counterparty exercise decisions are not considered in training, and training episodes continue past the expected exercise boundary. This is done to ensure that DRL agents experience prices below these boundaries, thereby improving robustness, as there is no guarantee that counterparties will exercise at the expected optimal boundary.

While all DRL agents in this study have the same architecture and hyperparameters, the training procedure differs across two rounds of experiments. The first series of experiments uses MC simulations of a GBM process to generate asset price data, and a second series of experiments uses MC simulations of a calibrated stochastic volatility model as the underlying asset price process.

### GBM Experiments: Training Procedures

In the GBM experiments, MC paths are simulated as:

$$S_{t+\Delta t} = S_t e^{\left(\mu - \frac{\sigma^2}{2}\right)\Delta t + \sigma * \sqrt{\Delta t} Z_t}, \tag{9}$$

where $Z_t \sim \mathcal{N}(0, 1)$ is a standard normal variable with mean zero and variance 1. Further, $\mu$ and $\sigma$ are the assets drift and volatility, which for GBM are both constant (Hull 2012). The discretization size is given by $\Delta t = \frac{T}{N}$, where $T$ is the time-to-maturity and $N$ is the number of discretization steps, which also represents the amount of rebalance points. In GBM experiments, the mean return is 5%, the volatility is 20%, and the time-to-maturity is one year.

Concerning option pricing, which is required for the agent reward, a binomial American put option price tree is constructed. The option price for an arbitrary asset price and time step is computed by interpolating the option tree. Constructing a binomial American put option price tree first requires the development of a binomial asset price tree. The asset price tree is constructed assuming a GBM process, where the asset price either moves up or down by a factor of $e^{\left(\mu - \frac{\sigma^2}{2}\right)\Delta t \pm \sigma * \sqrt{\Delta t} Z}$ (Hull 2012). The binomial asset price tree is used to compute the option payoffs at maturity, and option prices are then computed at each node by backward induction, i.e., computing the value function by comparing the value of holding the option to the value of exercise at each node of the tree.

### Stochastic Volatility Experiments: Training Procedures

In the second round of experiments, MC paths of a stochastic volatility model are simulated. The stochastic volatility model setup is given by:

$$\begin{aligned} \sigma_t &= \sigma_{t-1} + \nu \sigma_{t-1} \Delta B_t, \\ S_t &= S_{t-1} + \mu S_{t-1} \Delta t + \sigma_t S_{t-1} \Delta W_t, \end{aligned} \tag{10}$$

where $B_t$ and $W_t$ are Brownian motions with increments given by

$$\begin{bmatrix} \Delta W_t \\ \Delta B_t \end{bmatrix} = \sqrt{\Delta t} \begin{bmatrix} Z_1 \\ \rho Z_1 + \sqrt{1 - \rho^2} Z_2 \end{bmatrix}, \tag{11}$$

noting that $Z_1$ and $Z_2$ are independent standard normal random variables. In this stochastic volatility model, $\rho$ and $\nu$ are model coefficients that may be used to calibrate the model to empirical observations. This model is an adaptation of the SABR model given by Hagan et al. (2002). Hagan et al. (2002) present the model as

$$\sigma_t = \sigma_{t-1} + \nu \sigma_{t-1} \Delta B_t, \tag{12}$$

$$S_t = S_{t-1} + \sigma_t {S_{t-1}}^{\beta} \Delta W_t.$$

As such, for the model used in this study a non-zero drift term $\mu$ is added, and the $\beta$ parameter is set equal to 1.

Before calibrating the model to empirical option prices, a preliminary stochastic volatility experiment makes use of arbitrary model coefficients, setting $\rho$ to -0.4, which is reflective of the inverse relationship between asset prices and volatility (leverage effect) (Florescu and Pãsãricã 2009), and $\nu$ to 0.1. The stochastic volatility model is then calibrated to match true market assets. Specifically, given a listed price, maturity date, strike, and thereby an implied volatility, a DRL agent is implemented to hedge this specific option by using a calibrated stochastic volatility model for training. The empirical option data in this study is retrieved from the via the Bloomberg composite option monitor. For 8 symbols, put option prices on August 17th, 2023, for two different maturities (September 15th, 2023, and November 17th, 2023) and five different strikes were retrieved, giving a total of 80 options. Exhibit 1 summarizes the option data. Note that for each of the 80 options, the implied volatility is listed on the option monitor, and the option price is available by computing a mid-price of the bid-ask spread.

**Exhibit 1: Option Data Summary**

| Symbol | 08/17/2023 Close Price | Maturity Date | Strikes |
|---|---|---|---|
| AAPL | $174.00 | 9/15/2023<br>11/17/2023 | $165, $170, $175, $180, $185 |
| JNJ | $174.01 | 9/15/2023<br>11/17/2023 | $165, $170, $175, $180, $185<br>$150, $160, $170, $180, $190 |
| MA | $392.62 | 9/15/2023<br>11/17/2023 | $385, $390, $395, $400, $405 |
| META | $285.09 | 9/15/2023<br>11/17/2023 | $275, $280, $285, $290, $295 |
| MSFT | $316.88 | 9/15/2023<br>11/17/2023 | $305, $310, $315, $320, $325 |
| NKE | $105.05 | 9/15/2023<br>11/17/2023 | $97.5 $100, $105, $110, $115<br>$95, $97.5, $100, $105, $110 |
| NVDA | $433.44 | 9/15/2023<br>11/17/2023 | $425, $430, $435, $440, $445 |
| SPY | $436.29 | 9/15/2023<br>11/17/2023 | $434, $435, $436, $437, $438 |

For each option, model coefficients $\rho$ and $\nu$ are calibrated by setting the initial volatility to the retrieved implied volatility and minimizing the difference between the option mid-price and the average payoff given by 10000 model generated paths. Specifically, the constrained trust-region optimization algorithm is used. Given a parameter set for each symbol, the model parameters for each symbol are computed by taking an average. Note that the retrieved options are European, so no early exercise is considered when calibrating the model coefficients in this experiment. While the training and testing of the DRL agent considers American style options, model coefficients are calibrated without the exercise boundary, as the construction of said exercise boundary requires calibrated coefficients.

As for option pricing at each time step, recall that GBM experiments use the interpolation of a binomial option tree. However, tree models increase in complexity when considering a stochastic volatility model with two random variables, as the tree now requires a second spatial dimension. Without a binomial tree, American put option prices in this study are computed by leveraging a technique presented by Glau, Mahlstedt, and Potz (2018) that involves approximating the value function at each time-step by a Chebyshev polynomial approximation. Chebyshev approximation involves fitting a series of orthogonal polynomials to a function (Chebyshev 1864). A main advantage to using Chebyshev pricing method for this work is the increased efficiency over a simulation based approach. For example, given an exercise boundary, such as one generated the popular Longstaff-Schwartz Monte Carlo (LSMC) (Longstaff and Schwartz 2001) method, option prices for a given asset price and timestep may be computed by simulating several thousand MC paths of the underlying process to expiry or exercise, and the option price is computed as the average payoff. However, in training an RL agent, simulation-based pricing would add a considerable amount of time to the process, as a new series of simulations is required for each step of training. In this study, with 5000 episodes of 25 steps each, 125,000 sets of simulations would be required, an unnecessary addition to the already time-consuming process of training a neural network. Moreover, it is noted that this Chebyshev method is agnostic to the underlying asset evolution process.

Chebyshev interpolation involves weighing a set of orthogonal interpolating polynomials of increasing order so that their weighted sum approximates a known function between a pre-determined upper and lower bound at each time step, which are determined from the extremal excursions of the underlying asset paths. The general Chebyshev process for pricing American options first requires discretizing the computational space from $t_0 = 0$ to $t_N = T$. Then, the following three-step process ensues:

1. Starting one step before maturity at $t_{N-1}$, generate Chebyshev nodes between the upper and lower bounds and simulate paths from these nodes at $t_{N-1}$ to the maturity time $t_N$. Compute the continuation value for all nodes at $t_{N-1}$ as the discounted average payoff from all paths. Compute the value function for all Chebyshev nodes at $t_{N-1}$ by comparing the continuation value with the immediate exercise payoff.

2. For the next earliest time step, $t_{N-2}$, simulate the process from $t_{N-2}$ to $t_{N-1}$. To compute the discounted average payoffs from paths simulated from nodes at $t_{N-2}$, perform Chebyshev interpolation between the Chebyshev nodes at $t_{N-1}$, as the values at $t_{N-1}$ are known from step 1. Compute the value function for all nodes at time step $t_{N-2}$. Continue moving backward in time until $t = t_0$. By indexing the nodes where exercise was optimal, a boundary is computed.

3. Given the completion of steps 1 and 2, value functions are available at all nodes across the discretized computational space. As such, given an asset price level and a time step, one may use Chebyshev interpolation to compute an option price. Note that this step may be called at any time step in the DRL agent training process, without repeating steps 1 and 2.

To further exemplify why this Chebyshev method was chosen over LSMC, the reader is directed to Glau, Mahlstedt, and Potz (2018) for a full performance comparison of the Chebyshev and LSMC exercise boundaries. While the Chebyshev method is used for stochastic volatility experiments, the accuracy of the Chebyshev pricing method may be assessed by plotting the Chebyshev exercise boundary for a GBM case where there is a known, true boundary generated by a binomial tree with 5000 nodes. Specifically, consider an at-the-money American put option struck at $100 with a 1-year maturity. Letting the underlying asset follow a GBM process, a comparison of the Chebyshev and binomial boundaries is given in Exhibit 2, and the results show a pronounced agreement between the two methods.

**Exhibit 2: Exercise Boundary Comparison**

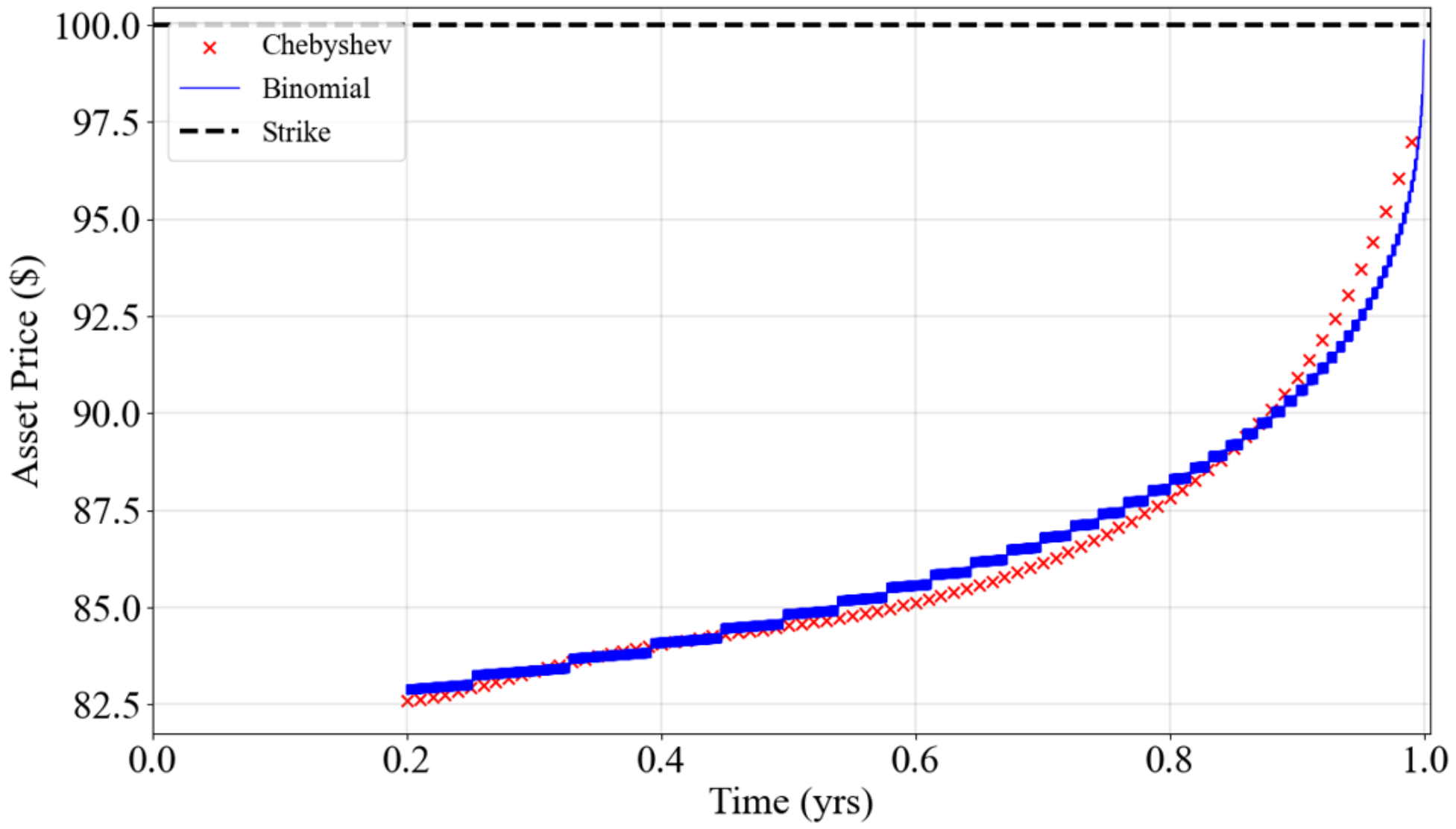

## TESTING PROCEDURES

Once a DRL agent is trained, a testing procedure is designed to assess the hedging performance of the agent. First, note that the final P&L of the hedging agent is computed by tracking a money-market account position, denoted as $B$, through each time step until maturity:

$$t = 0: B_0 = C_0 - S_0 A_0,$$

$$t = n: B_n = B_{(n-1)} e^{r\Delta t} - (A_n - A_{n-1}) S_n - \lambda |A_{n-1} - A_n| S_n, \tag{13}$$

$$Exercise, t = T^*: Final\ P\&L = \ B_{T*} = B_{(T*-\Delta t)} e^{r\Delta t} + S_{T*} A_{T*-\Delta t} - (K - S_{T*})_+$$

Given this defined performance metric, the main requirements for agent testing are:

1. The generation of asset price data, which is required for the DRL agent state and the P&L calculations.

2. The American put option price at the initial time step, used in the P&L calculation.
3. The construction of an exercise boundary to be used by the counterparty.
4. The design of benchmark hedging strategies for comparison.

Once again, this section will be grouped into GBM and stochastic volatility model experiments.

**GBM Experiments: Testing Procedure**

As in training, GBM tests use MC paths of the GBM process for testing. Moreover, given that an American option tree was computed for training, the tree is used to generate the initial option price in testing. Further, by indexing the nodes at which exercise is optimal, an exercise boundary to be used by the counterparty is constructed. Given that binomial trees are constructed for both the asset and option price processes, this study uses a binomial tree hedge as a testing benchmark for the DRL agent. At a given timestep, the binomial tree hedge for an American put option, which in this study is denoted $\beta_t$, is given by

$$\beta_t = -\frac{C_t u - C_t d}{S_t u - S_t d}. \tag{14}$$

$C_t u$ and $C_t d$ are the option prices at the next time step at the up and down nodes, respectively, and $S_t u$ and $S_t d$ are the asset prices at the next time step at the up and down nodes, respectively (Hull 2012). The binomial hedge may be interpolated from the hedge tree given an arbitrary GBM asset price and an interpolated option price.

A BS Delta hedge is also used as a comparative benchmark for the DRL agent. While the BS model is derived using European options (Hull 2012), a Delta hedge still provides a relevant comparator. For a given asset price and time-to-maturity, the Delta of a put option is given as

$$\Delta_t = N\left(\frac{\log\left(\frac{S_t}{K}\right) + \left(r + \frac{\sigma^2}{2}\right)(T-t)}{\sigma\sqrt{T-t}}\right) - 1, \tag{12}$$

where $N(*)$ is the cumulative normal distribution function, $r$ is the risk-free rate (5% for all experiments), and $K$ is the strike price (Hull 2012). All GBM tests are conducted using 10000 GBM sample paths with 100 rebalance points each, and the initial asset price and strike price are both set at $100.

**Stochastic Volatility Experiments: Testing Procedure**

Recall that to establish a baseline, the first stochastic volatility experiment trains the DRL agent using arbitrary model coefficients. The option maturity is set to one month, and the DRL agent trained with paths from said stochastic volatility model with arbitrary coefficients is tested across 10000 episodes with 21 rebalance steps each (daily rebalancing), noting that the initial volatility is 20%, and the initial asset price and strike price are set to $100.

Upon completion of this experiment, the market calibrated DRL agents are tested by simulating 10000 asset paths using the respective calibrated model, the implied initial volatility, and the option maturity. The calibrated Chebyshev exercise boundary is used for determining counterparty decisions. Daily rebalancing is assumed, and therefore the number of rebalance points in each testing episode is the number of trading days between August $17^{th}$, 2023, and the option maturity date. To give insight into how a DRL agent trained with a market-calibrated stochastic volatility model will perform in practice, a final experiment in this study evaluates the performance of each of the 80 DRL agents on the true asset path that was realized between August 17th, 2023, and the maturity date. The price data between August 17th, 2023, and November 17th, 2023, for all 8 symbols used were retrieved from Yahoo finance and are listed in the Appendix. As for the design of a comparative benchmark strategy, only the BS Delta is used as a comparative benchmark as a binomial tree is not constructed for stochastic volatility experiments. The updated volatility at each time step is substituted into the BS Delta calculation.

## RESULTS

This section presents and discusses the results of numerical experiments. As described in the methodology, a DRL agent is first trained using simulations of a GBM process, and the testing results of this GBM trained agent are presented first. Not only are GBM experiments conducted with and without transaction costs, the robustness of the DRL agent is assessed by conducting an experiment at a volatility level higher than what was seen by the DRL agent in training. Next, the real-world applicability is assessed by testing DRL agents trained with market-calibrated stochastic volatility models, noting that the first stochastic volatility test uses arbitrary parameters.

### Geometric Brownian Motion Experiment

In the first experiment, a DRL agent is trained using GBM asset paths. Exhibits 3 and 4 show the resulting final P&L distributions when the transaction cost rates, $\lambda$, are 0 and 3%, respectively. Note that separate subplots are used for ease of comparison, and the DRL agent results are the same when compared to the BS Delta and binomial strategies in each subplot. Exhibit 5 summarizes the mean and standard deviations of the final P&L for DRL, BS Delta, and binomial hedging, and the highest (lowest) mean (standard deviation) is italicized and underlined for each transaction cost rate. Exhibits 6 and 7 show sample hedging actions for one asset path that finishes out of the money (OTM), and one asset path that crosses the early exercise boundary.

**Exhibit 3: Final P&L Distributions Across 10000 GBM Paths (λ = 0%)**

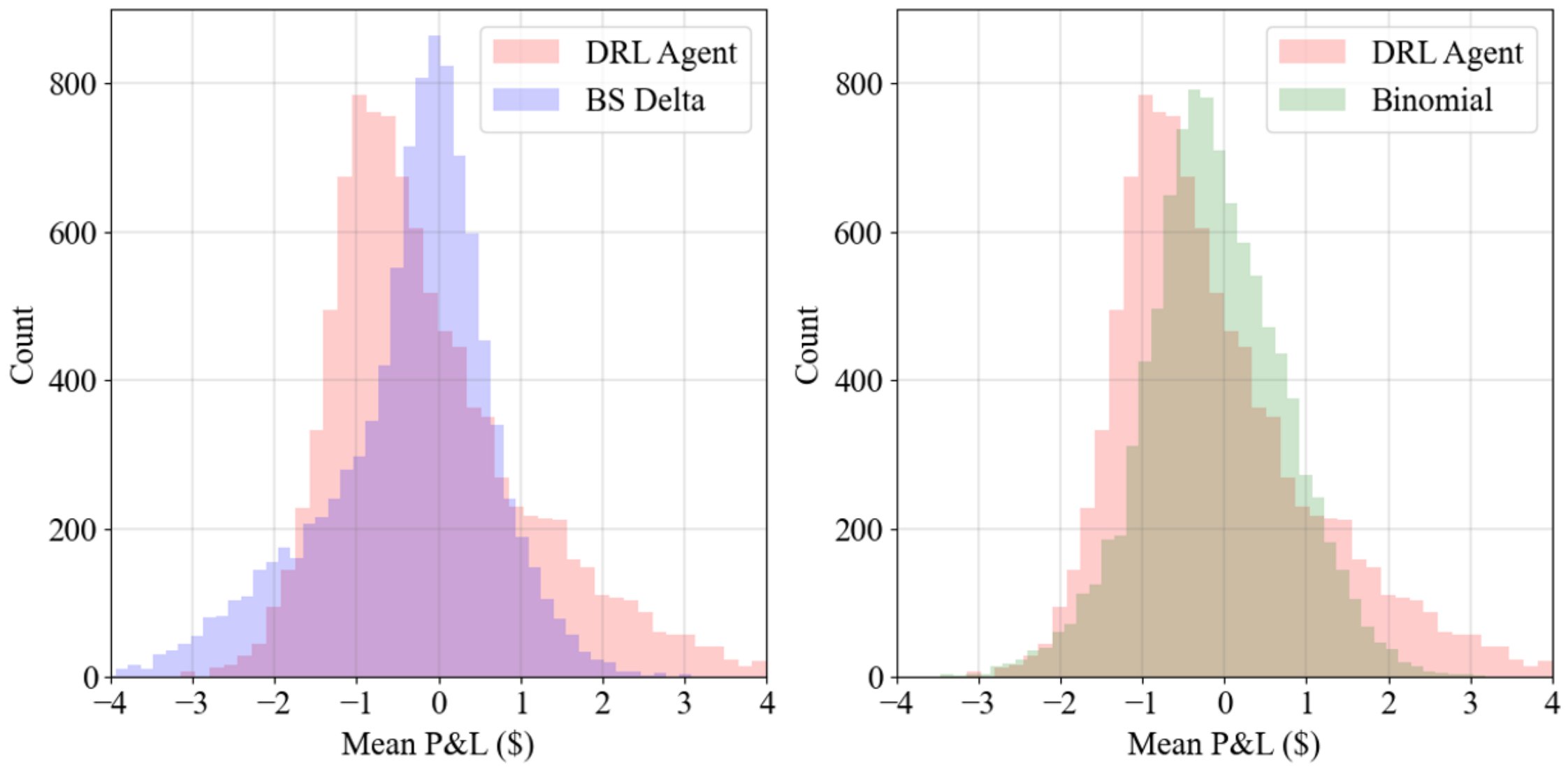

**Exhibit 4: Final P&L Distributions Across 10000 GBM Paths (λ = 3%)**

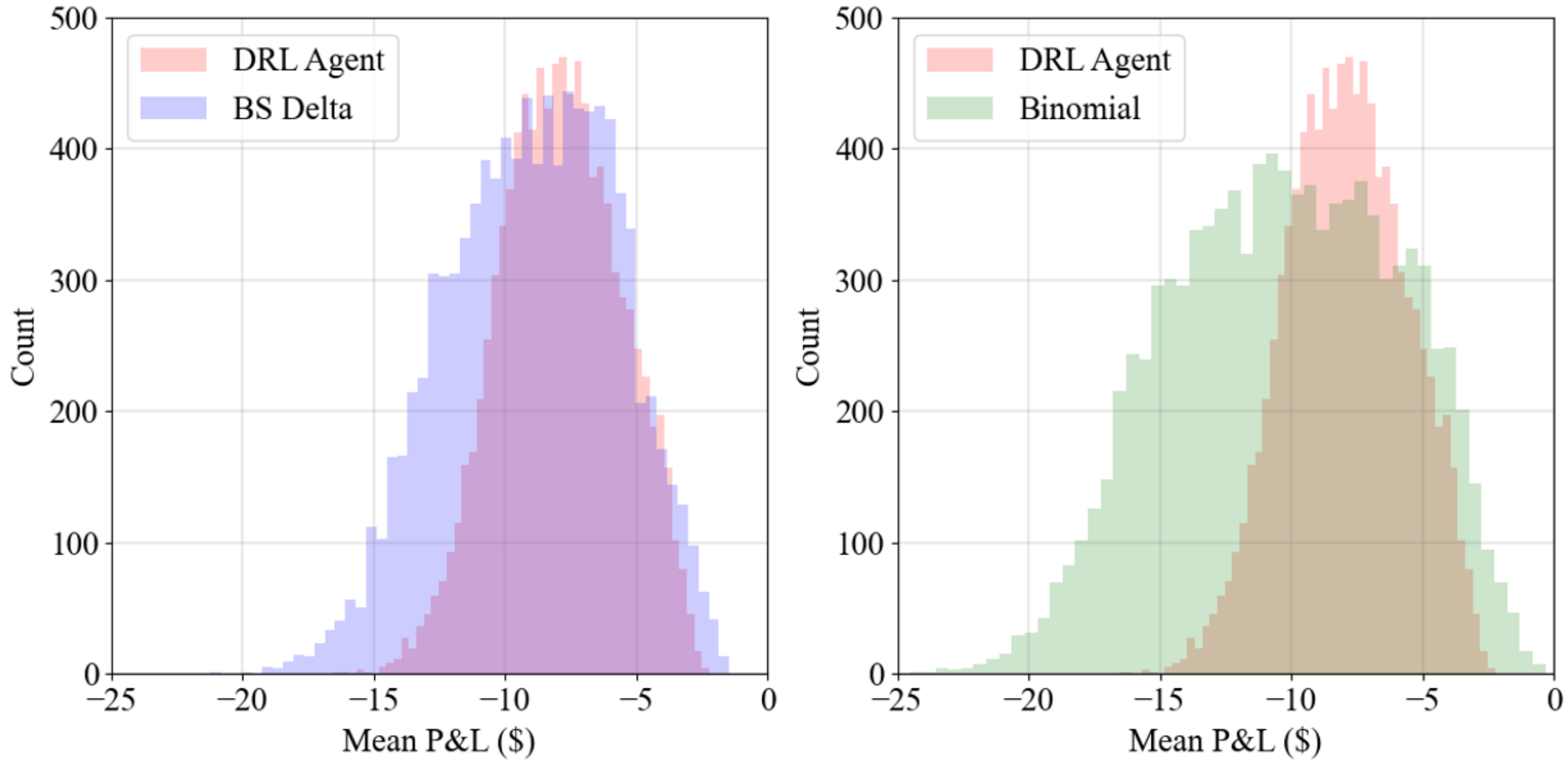

**Exhibit 5: Final P&L Statistics Across 10000 GBM Paths**

| | λ = 0% | | λ= 3% | |
|---|---|---|---|---|
| | **Mean** | **Std. Dev** | **Mean** | **Std. Dev** |
| **DRL Agent** | *-$0.10* | $1.20 | *-$7.90* | *$2.30* |
| **BS Delta** | -$0.36 | $1.02 | -$8.97 | $3.24 |
| **Binomial** | -$0.12 | *$0.86* | -$10.26 | $4.33 |

**Exhibit 6: Sample Hedging Actions for 1 GBM Path (No Early Exercise)**

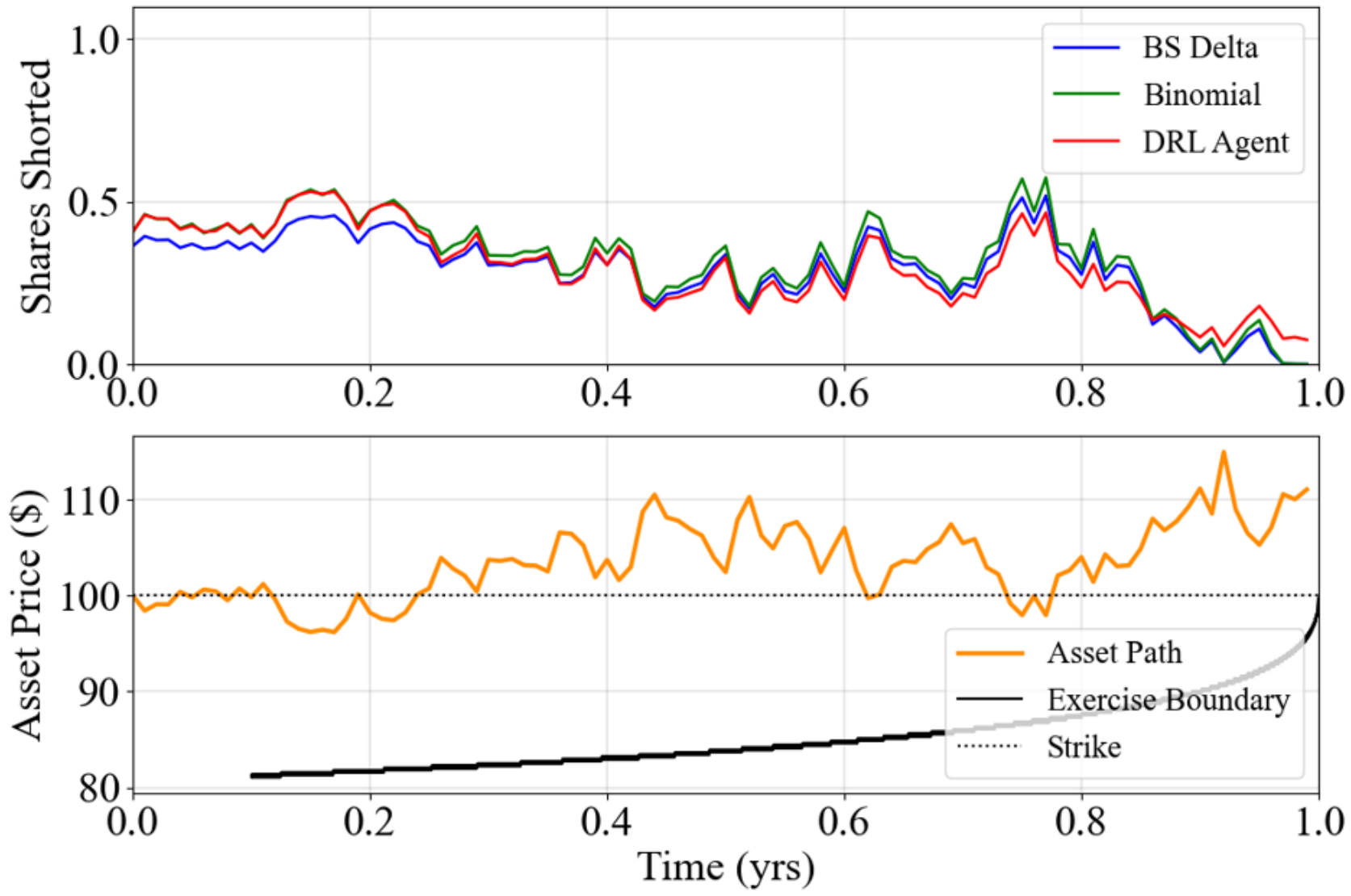

**Exhibit 7: Sample Hedging Actions for 1 GBM Path (Early Exercise)**

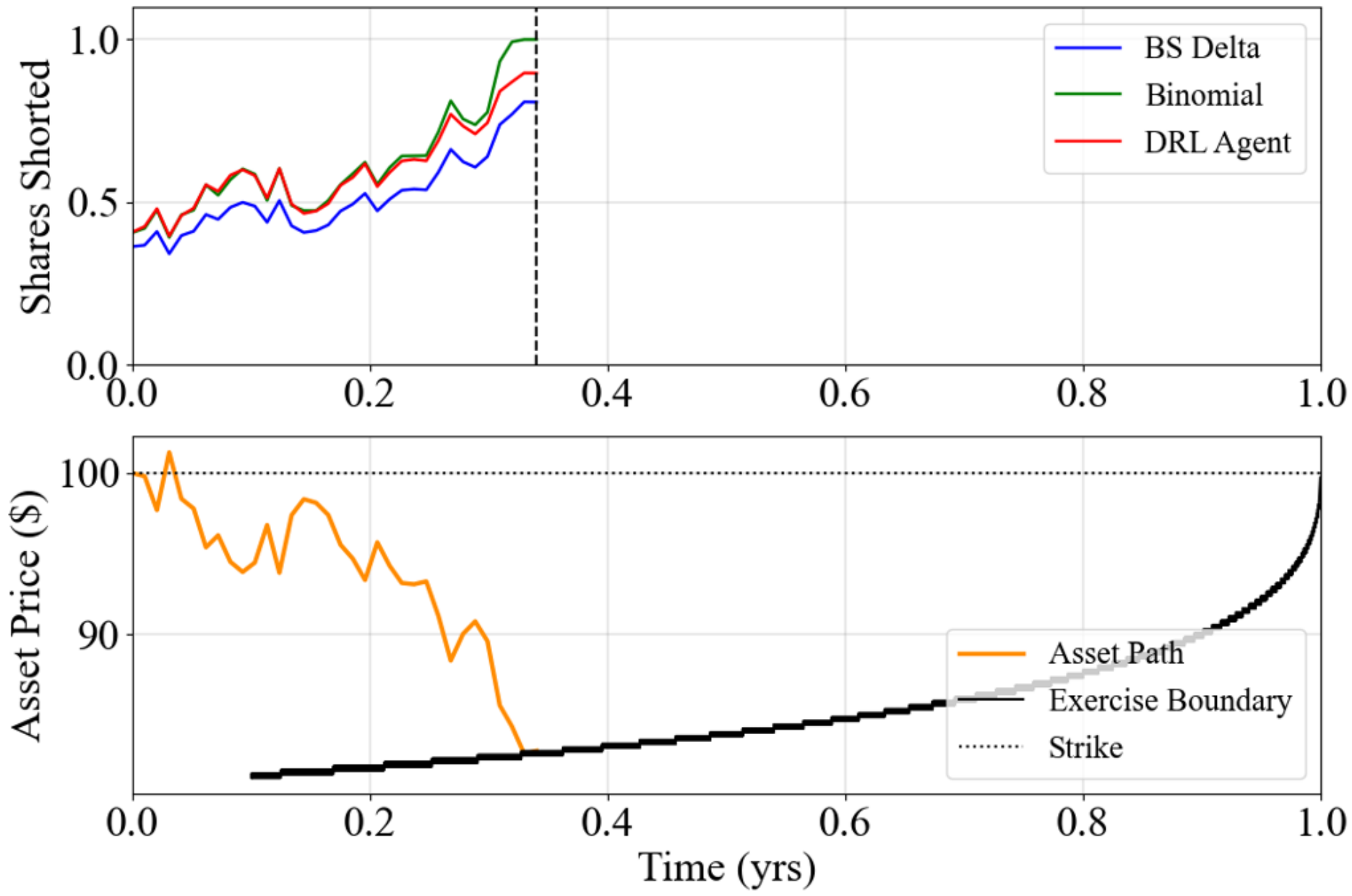

A comparison of Exhibits 3 and 4 shows the impact of a cost conscious DRL agent. When there are no transaction costs, the DRL agent hedges effectively, producing a higher mean final P&L than the BS Delta method. This is an expected result, as the BS Delta method does not consider early exercise. However, the DRL agent does not achieve a higher mean final P&L than the binomial strategy, and the standard deviation of final P&L's is highest for the DRL agent. However, when there are 3% transaction costs, the DRL agent achieves a higher mean final P&L and lower standard deviation than both the BS Delta and binomial strategies. Exhibits 5 and 6 aid in illustrating the DRL agent's awareness of premature exercise risk while maintaining a focus on transaction costs. At early time steps, the DRL agent hedges similarly to the binomial strategy, and the BS Delta strategy does not consider early exercise and, therefore, has the lowest hedge position. While the three strategies converge when the threat of early exercise decreases and the asset price moves out of the money, as is in the case in Exhibit 6, the DRL agent tends to react more slowly to abrupt price movements. When there is a spike in the required asset position, the DRL agent tends to under-hedge relative to the other two strategies. Likewise, the DRL agent tends to over-hedge when there is a dip in the required asset position. This displays the cost-consciousness of the DRL agent and helps show why the DRL method outperforms the other two strategies when transaction costs are present.

The second GBM experiment aims to display the robustness of the DRL agent. Call to mind that the DRL agent is trained with a volatility of 20%, the binomial strategy is computed with trees constructed with 20% volatility, and the BS Delta strategy uses a volatility of 20% to compute both the option price and Delta. As such, let this 20% volatility be labelled as $\sigma_{agent}$. Now consider a scenario wherein GBM asset paths evolve with a volatility of 24%. Moreover, let the option buyer use the correct volatility of 24% to compute an exercise boundary for exercise decisions. As such, this volatility of 24% is given the label $\sigma_{buyer}$. In this scenario, not only are the hedging agents using an incorrect volatility to hedge the option, but the initial option price is computed with a lower volatility, and the option is therefore sold under value. Exhibits 8 through 10 summarize the results of this volatility experiment.

**Exhibit 8: Final P&L Distributions for 10000 GBM Paths ($\sigma^{agent}$= 20%, $\sigma^{buyer}$ = 24%, λ=0%)**

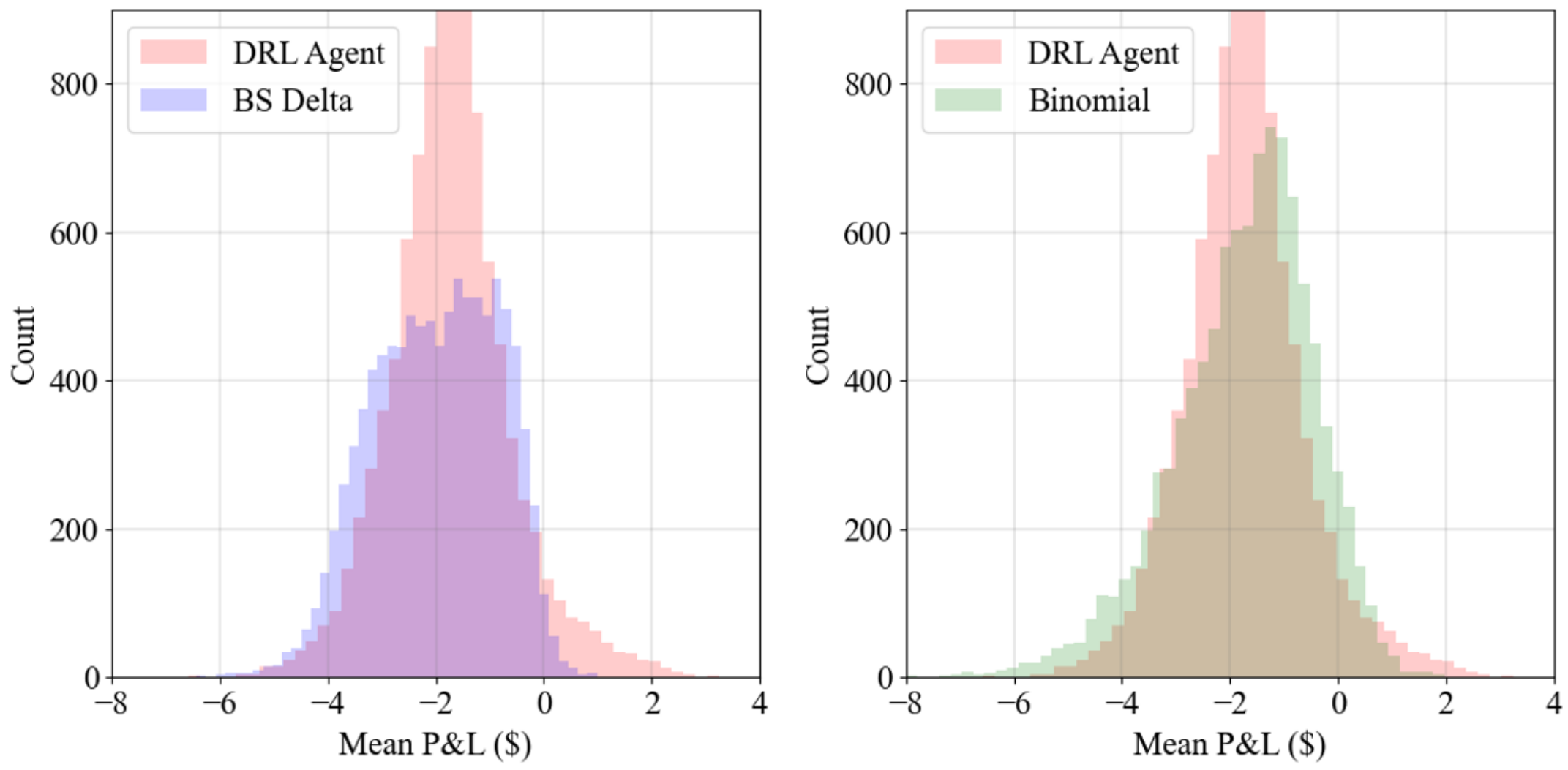

**Exhibit 9: Final P&L Distributions for 10000 GBM Paths ($\sigma^{agent}$= 20%, $\sigma^{buyer}$ = 24%, λ=3%)**

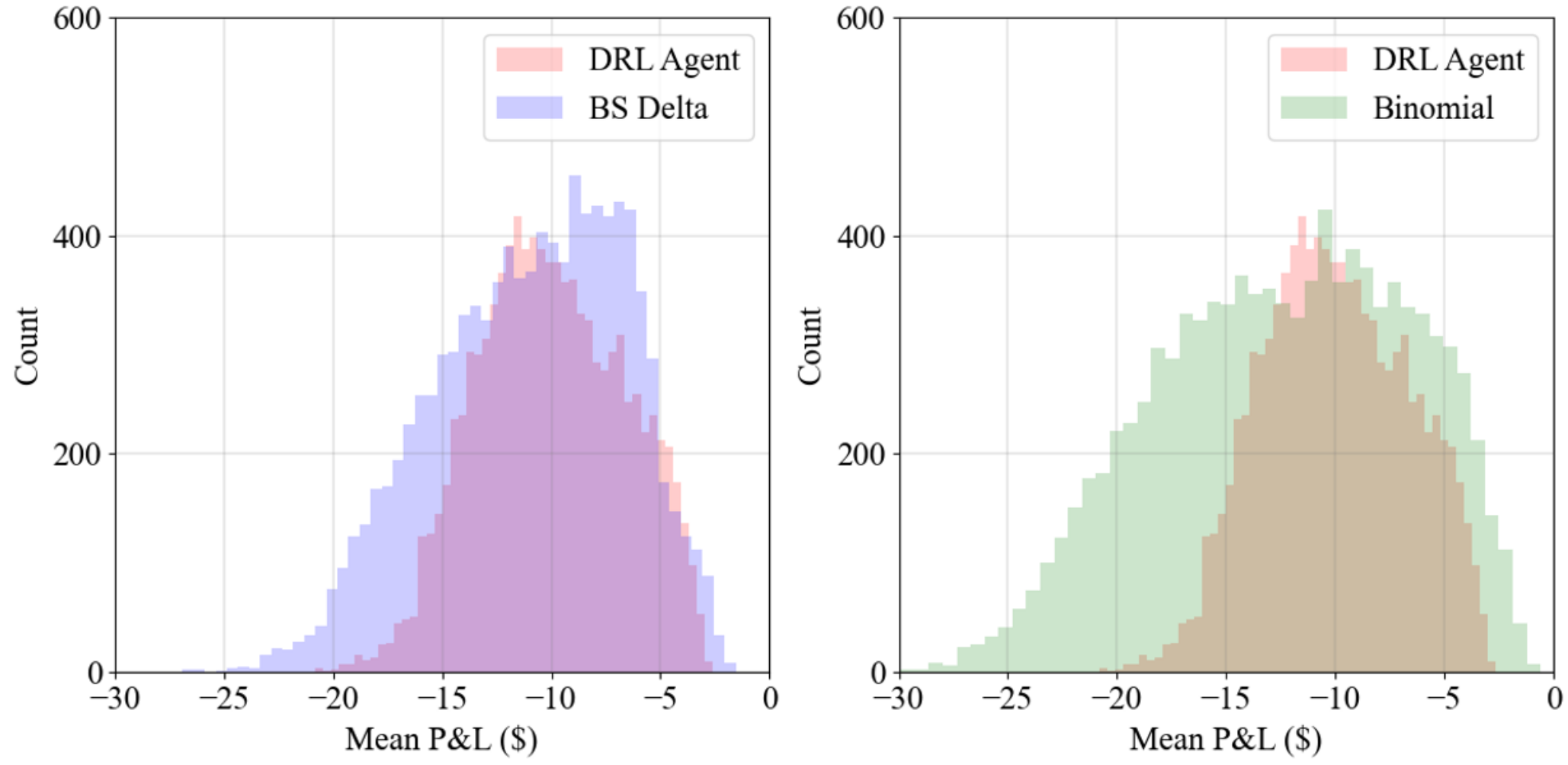

| **Exhibit 10: Final P&L Statistics Across 10000 GBM Paths for $\sigma_{agent}$ = 20%, $\sigma_{buyer}$ = 24%** | | | | |
|---|---|---|---|---|
| | **λ = 0%** | | **λ = 3%** | |
| | **Mean** | **Std. Dev** | **Mean** | **Std. Dev** |
| **DRL Agent** | *-$1.72* | *$1.11* | *-$10.02* | *$3.35* |
| **BS Delta** | -$1.99 | $1.16 | -$11.00 | $4.39 |
| **Binomial** | -$1.73 | $1.31 | -$12.42 | $5.79 |

The results of Exhibits 8 through 10 show that the DRL agent is more robust to an inaccurate volatility estimate than the BS Delta and binomial tree strategies. With no transaction costs, the DRL agent has a higher mean final P&L than the other two strategies, and a lower standard deviation than the binomial strategy. With transaction costs considered, the DRL agent once again outperforms both the BS Delta and the binomial strategies, achieving a higher mean final P&L, as well as a lower standard deviation.

### SABR Experiments

To assess a real-world application of DRL hedging, agents are trained using market-calibrated stochastic volatility model paths. Recall however that to establish a baseline, a first experiment uses arbitrary model coefficients. For this initial experiment, Exhibits 11 and 12 show the final P&L distributions of the DRL and BS Delta strategies under 0% and 3% transaction costs, respectively, and Exhibit 13 shows the summary statistics of both cases.

**Exhibit 11: Final P&L Distributions for 10000 SABR Paths for ρ = -0.4, υ = 0.1 (λ=0%)**

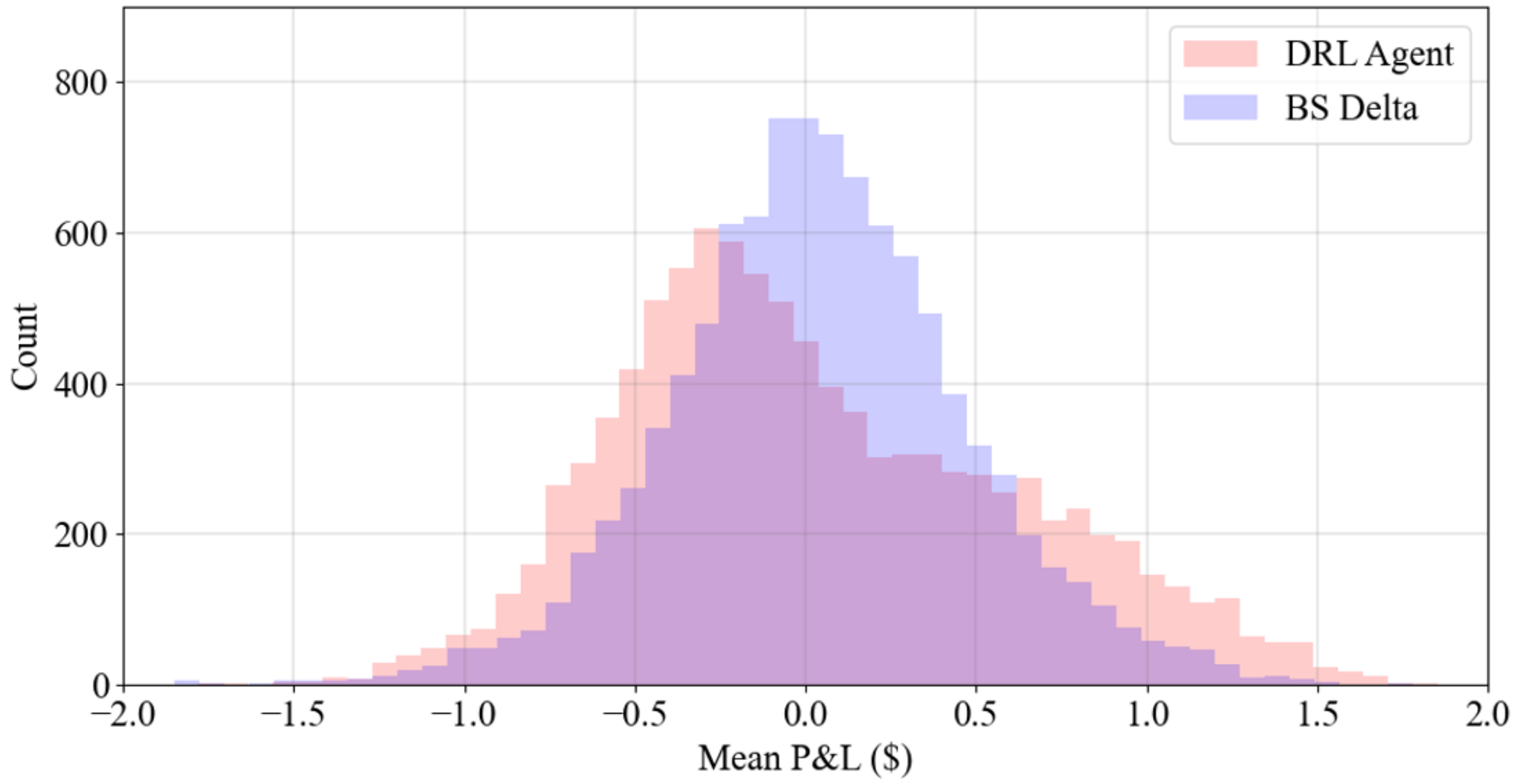

**Exhibit 12: Final P&L Distributions for 10000 SABR Paths for ρ = -0.4, υ = 0.1 (λ=3%)**

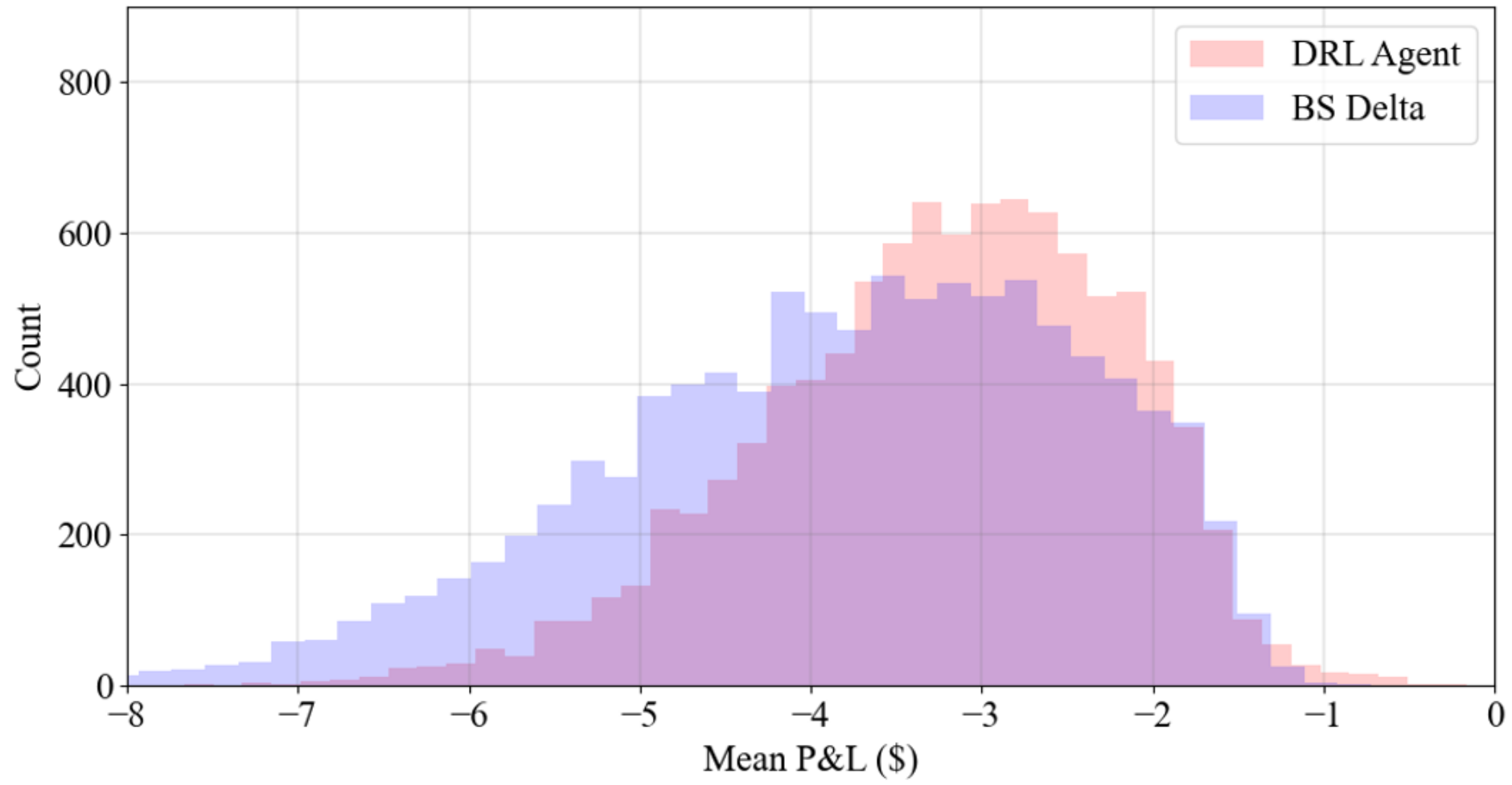

**Exhibit 13: Final P&L Statistics Across 10000 Stochastic Volatility Paths for $\rho$ = -0.4 and $\nu$ = 0.1**

| | TC = 0% | | TC = 3% | |
|---|---|---|---|---|
| | **Mean** | **Std. Dev** | **Mean** | **Std. Dev** |
| **DRL Agent** | $0.035 | $0.58 | *-$3.23* | *$1.05* |
| **BS Delta** | *$0.042* | *$0.44* | -$3.79 | $1.39 |

The results are consistent with the GBM experiments, showing that under transaction costs, the DRL agent outperforms the BS Delta strategy, producing a higher mean final P&L and a lower standard deviation. The real-world application may now be examined using the DRL agents trained with market calibrated models. To present results without listing all 80 options, the mean and standard deviation of mean final P&L result across five strikes for each symbol and maturity date is computed. These final P&L statistics for both the DRL agent and the BS Delta strategy are summarized in Exhibit 14 for a transaction cost-free scenario, and in Exhibit 15 for a 3% transaction costs. The reader should not compare results between different rows, as the results are functions of both symbol prices and maturities. The DRL and BS Delta final P&Ls statistics for all 80 options are listed in the Appendix.

**Exhibit 14: Final Mean P&L Statistics Across 5 Strikes using 10000 Market Calibrated Stochastic Volatility Paths (λ = 0%)**

| Symbol | Maturity Date | Mean of DRL Mean | Mean of Delta Mean | Mean of DRL SD | Mean of Delta SD |
|---|---|---|---|---|---|
| AAPL | 9/15/2023 | *$0.06* | -$0.01 | $1.29 | *$0.77* |
| | 11/17/2023 | *$0.02* | -$0.15 | $2.23 | *$1.49* |
| JNJ | 9/15/2023 | *$0.07* | $0.00 | $1.46 | *$0.80* |
| | 11/17/2023 | *$0.01* | -$0.27 | $1.94 | *$1.07* |
| MA | 9/15/2023 | *$0.21* | $0.04 | $3.49 | *$1.72* |
| | 11/17/2023 | *$0.02* | -$0.28 | $4.05 | *$3.02* |
| META | 9/15/2023 | *$0.26* | $0.17 | $3.25 | *$2.23* |
| | 11/17/2023 | *$0.18* | -$0.06 | $5.33 | *$3.95* |
| MSFT | 9/15/2023 | *$0.15* | $0.07 | $2.62 | *$1.81* |
| | 11/17/2023 | *$0.03* | -$0.25 | $3.94 | *$3.13* |
| NKE | 9/15/2023 | *$0.04* | -$0.03 | $0.99 | *$0.50* |
| | 11/17/2023 | *$0.00* | -$0.09 | $1.18 | *$0.79* |
| NVDA | 9/15/2023 | *$0.66* | $0.58 | $7.05 | *$6.33* |
| | 11/17/2023 | *-$0.05* | -$0.10 | $10.77 | *$7.34* |
| SPY | 9/15/2023 | *$0.12* | $0.03 | $2.42 | *$1.54* |
| | 11/17/2023 | *-$0.01* | -$0.33 | $3.13 | *$2.35* |

**Exhibit 15: Final Mean P&L Statistics Across 5 Strikes using 10000 Market Calibrated Stochastic Volatility Paths (λ = 3%)**

| Symbol | Maturity Date | Mean of RL Mean | Mean of Delta Mean | Mean of RL SD | Mean of Delta SD |
|---|---|---|---|---|---|
| AAPL | 9/15/2023 | *-$4.26* | -$5.37 | *$2.26* | $2.65 |
| | 11/17/2023 | *-$9.33* | -$11.44 | *$4.61* | $5.09 |
| JNJ | 9/15/2023 | *-$4.14* | -$5.41 | *$2.52* | $2.71 |
| | 11/17/2023 | *-$6.57* | -$7.83 | *$4.85* | $5.30 |
| MA | 9/15/2023 | *-$10.48* | -$14.07 | $6.47 | *$5.70* |
| | 11/17/2023 | *-$25.57* | -$26.70 | *$11.03* | $11.19 |
| META | 9/15/2023 | *-$9.14* | -$10.25 | $5.00 | *$4.55* |
| | 11/17/2023 | *-$17.25* | -$19.47 | $9.24 | *$8.69* |
| MSFT | 9/15/2023 | *-$9.74* | -$11.32 | *$4.39* | $4.78 |
| | 11/17/2023 | *-$19.63* | -$21.77 | *$7.96* | $9.00 |
| NKE | 9/15/2023 | *-$2.08* | -$2.79 | *$1.54* | $1.66 |
| | 11/17/2023 | *-$5.19* | -$6.16 | *$2.59* | $3.13 |
| NVDA | 9/15/2023 | *-$15.74* | -$15.92 | $10.01 | *$8.77* |
| | 11/17/2023 | -$32.26 | *-$30.55* | $16.55 | *$13.37* |
| SPY | 9/15/2023 | *-$13.84* | -$16.36 | *$4.89* | $5.91 |
| | 11/17/2023 | *-$27.95* | -$29.12 | *$10.56* | $12.11 |

When there are no transaction costs, both the DRL agent and the BS Delta strategy achieve near-zero means in all cases, indicating effective hedging performance. These results show that the DRL strategy achieves a higher mean final P&L in all 16 cases, while the BS Delta strategy yields a lower standard deviation in all cases. With a 3% transaction cost rate, the results indicate that the DRL agent achieves a higher mean of mean final P&Ls across the five strikes than the BS Delta strategy in all but one symbol and maturity date combinations (NVDA 11/17/2023). Moreover, the DRL agent achieves a lower mean standard deviation in 11 of 16 cases, exemplifying that the DRL agents hedge in a consistent manner that achieves a lower variance on average than a BS Delta strategy. This outcome implies that when transaction costs are incorporated, training a DRL agent with calibrated stochastic volatility model parameters is more effective than computing a BS Delta hedge using the model-derived volatility at each time step.

Now, the DRL agent is tested on the empirical asset price data between the sale and maturity dates. Note that as the experiment aims to assess only the performance of the DRL agent in a realistic scenario, only an environment with a 3% transaction cost rate is used for testing. Exhibit 16 summarizes the mean final P&L for both the DRL agent and the BS Delta strategy across the five strikes for each symbol and maturity. Again, results should not be compared across rows, as the results are a function of both the underlying asset price process and the maturity date. The final hedging P&L for all 80 options using the true asset paths are listed in the Appendix.

**Exhibit 16: Mean Final P&L's Across 5 Strikes using Empirical Asset Paths**

| Symbol | Maturity Date | RL Mean | BS Delta Mean |
|---|---|---|---|
| AAPL | 9/15/2023 | *-$5.04* | -$7.97 |
| | 11/17/2023 | *-$9.64* | -$12.14 |
| JNJ | 9/15/2023 | *-$2.67* | -$2.74 |
| | 11/17/2023 | *-$2.32* | -$3.33 |
| META | 11/17/2023 | -$6.98 | *-$6.97* |
| | 9/15/2023 | *-$16.15* | -$18.55 |
| MA | 9/15/2023 | -$6.28 | -$6.13 |
| | 11/17/2023 | -$15.12 | *-$11.92* |
| MSFT | 9/15/2023 | *-$4.86* | -$6.28 |
| | 11/17/2023 | -$15.56 | *-$15.51* |
| NKE | 11/17/2023 | *-$1.18* | -$1.97 |
| | 9/15/2023 | *-$5.50* | -$6.28 |
| NVDA | 9/15/2023 | *$6.22* | $5.30 |
| | 11/17/2023 | *-$16.07* | -$17.24 |
| SPY | 9/15/2023 | *-$9.98* | -$10.91 |
| | 11/17/2023 | *-$29.26* | -$34.75 |

Using the actual asset paths for testing, the DRL agent outperforms the BS Delta strategy across the five strikes in 12 of 16 instances. As such, this outcome shows that not only do DRL agents achieve strong performance under simulated asset paths, maintaining consistency with the stochastic volatility model employed during training, but DRL agents also possess a desirable robustness in effectively hedging against empirical stock price movements. This is an important result, as it shows that on any given day, observed option prices may be used to train a DRL agent capable of hedging said option. Consider, for example, a trader with a basket of 80 options: the results from this work suggest that this trader can train a DRL agent for each option and achieve better performance than a BS Delta strategy. As such, a worthwhile future direction for this research would be to train the DRL agents anew on each day given the observed option prices. Further, as DRL becomes more prevalent in the hedging space, a worthwhile next step could be a parametric study that evaluates the impacts of various training hyperparameters such as the learning rates, episodes, re-balance steps, batch sizes, and reward functions. As there is currently no consensus in the DRL hedging literature towards hyperparameters and training best practices, this future study would help provide key information for practitioners attempting to employ their own DRL agents.

## CONCLUSIONS

In summary, this article fills a void in the DRL hedging field, which primarily focuses on European options, by introducing the application of DRL agents to hedge American put options. The study utilizes the DDPG algorithm for this purpose. Notably, while the state variables—comprising the asset price, current holding, and time-to-maturity—conform to existing literature, a novelty of this research lies in a unique reward function. This reward function accounts for both the negative absolute difference between the option value and the underlying asset position value and incorporates a quadratic penalty for transaction costs. The first round of experiments uses GBM asset paths for both training and testing, and initial results show that under the presence of transaction costs, the DRL agent outperforms the BS Delta and binomial tree hedging strategies. Another preliminary result of this work is that when the hedging agent underestimates the volatility, the DRL agent once again outperforms the BS Delta and Binomial strategies.

This article extends the findings of the GBM results by training DRL agents using asset paths generated through a stochastic volatility model. Stochastic volatility model parameters are calibrated to each symbol by using 80 options from eight symbols, featuring two maturities and five strikes. In contrast to the GBM approach, where a binomial tree interpolation provides option prices at each time step, option prices for stochastic volatility experiments are computed using a Chebyshev interpolation method, which eliminates the requirement to compute average payoffs from MC simulations up to maturity or early exercise at each training step. The outcomes reveal that the DRL agent not only outperforms the BS Delta strategy when tested with paths generated from the same stochastic volatility model used in training, but the DRL agent also exhibits superior hedging performance, on average, as compared to BS Delta when testing with the realized underlying assets between the option sale date and maturity. Finally, note that as the Chebyshev pricing method is model agnostic, this work contributes a general framework that may be extended to train DRL agents for any desired underlying asset process.

These results imply that DRL agents may be employed in an empirical setting to effectively hedge American put options by calibrating the training process to listed option data. A recommended future direction for more optimal results would be to retrain the agent more frequently as new option data becomes available. Further, a future study that examines the effect of training hyperparameters is recommended to aid practitioners in starting to train their own DRL agents.

# Appendix A

| Exhibit A1: Final Mean P&L Statistics using 10000 Market Calibrated Stochastic Volatility Paths (λ = 0%) | | | | | | |
|---|---|---|---|---|---|---|
| **Symbol** | **Maturity** | **Strike** | **RL Mean** | **Delta Mean** | **RL STD** | **Delta STD** |
| AAPL | 9/15/2023 | $165.00 | $0.03 | -$0.03 | $1.09 | $0.78 |
| | | $170.00 | $0.06 | $0.04 | $1.31 | $0.86 |
| | | $175.00 | $0.08 | $0.04 | $1.12 | $0.87 |
| | | $180.00 | $0.05 | -$0.01 | $0.89 | $0.77 |
| | | $185.00 | $0.07 | -$0.09 | $2.05 | $0.59 |
| | 11/17/2023 | $165.00 | -$0.03 | -$0.20 | $2.28 | $1.45 |
| | | $170.00 | -$0.06 | -$0.15 | $2.18 | $1.59 |
| | | $175.00 | $0.02 | -$0.11 | $2.03 | $1.58 |
| | | $180.00 | $0.07 | -$0.08 | $1.71 | $1.52 |
| | | $185.00 | $0.08 | -$0.23 | $2.95 | $1.33 |
| JNJ | 9/15/2023 | $165.00 | $0.05 | -$0.02 | $1.25 | $0.71 |
| | | $170.00 | $0.01 | $0.01 | $1.56 | $0.77 |
| | | $175.00 | $0.04 | $0.04 | $1.11 | $0.80 |
| | | $180.00 | $0.10 | $0.00 | $1.25 | $0.83 |
| | | $185.00 | $0.13 | -$0.04 | $2.14 | $0.90 |
| | 11/17/2023 | $150.00 | -$0.02 | -$0.28 | $2.03 | $1.07 |
| | | $160.00 | -$0.03 | -$0.24 | $2.10 | $1.11 |
| | | $170.00 | $0.01 | -$0.18 | $1.61 | $1.38 |
| | | $180.00 | $0.16 | -$0.04 | $2.70 | $0.94 |
| | | $190.00 | -$0.07 | -$0.62 | $1.28 | $0.84 |
| MA | 9/15/2023 | $385.00 | $0.13 | $0.06 | $2.48 | $1.78 |
| | | $390.00 | $0.11 | $0.09 | $2.79 | $1.80 |
| | | $395.00 | $0.25 | $0.04 | $3.40 | $1.77 |
| | | $400.00 | $0.24 | $0.04 | $2.89 | $1.69 |
| | | $405.00 | $0.32 | -$0.02 | $5.87 | $1.56 |
| | 11/17/2023 | $385.00 | $0.04 | -$0.25 | $3.65 | $2.93 |
| | | $390.00 | -$0.23 | -$0.39 | $5.48 | $3.19 |
| | | $395.00 | $0.06 | -$0.30 | $4.17 | $3.20 |
| | | $400.00 | $0.03 | -$0.17 | $3.25 | $2.68 |
| | | $405.00 | $0.18 | -$0.28 | $3.72 | $3.10 |
| META | 9/15/2023 | $275.00 | $0.15 | $0.12 | $3.23 | $2.18 |
| | | $280.00 | $0.19 | $0.15 | $3.18 | $2.23 |
| | | $285.00 | $0.37 | $0.19 | $3.29 | $2.29 |
| | | $290.00 | $0.18 | $0.17 | $2.62 | $2.28 |
| | | $295.00 | $0.41 | $0.21 | $3.91 | $2.18 |
| | 11/17/2023 | $275.00 | $0.17 | -$0.16 | $4.55 | $3.97 |
| | | $280.00 | $0.00 | -$0.22 | $6.07 | $4.11 |
| | | $285.00 | $0.37 | $0.17 | $5.95 | $3.70 |
| | | $290.00 | $0.20 | -$0.12 | $4.71 | $4.12 |
| | | $295.00 | $0.16 | $0.04 | $5.35 | $3.85 |
| MSFT | 9/15/2023 | $305.00 | $0.14 | $0.04 | $3.53 | $1.74 |
| | | $310.00 | $0.11 | $0.04 | $2.05 | $1.83 |
| | | $315.00 | $0.08 | $0.09 | $2.98 | $1.85 |
| | | $320.00 | $0.23 | $0.10 | $2.09 | $1.84 |
| | | $325.00 | $0.18 | $0.09 | $2.45 | $1.78 |
| | 11/17/2023 | $305.00 | $0.01 | -$0.24 | $4.15 | $2.98 |
| | | $310.00 | -$0.06 | -$0.27 | $4.24 | $3.23 |
| | | $315.00 | $0.11 | -$0.20 | $3.98 | $3.09 |
| | | $320.00 | $0.06 | -$0.22 | $3.83 | $3.18 |
| | | $325.00 | $0.01 | -$0.30 | $3.51 | $3.17 |
| NKE | 9/15/2023 | $97.50 | $0.04 | -$0.01 | $0.78 | $0.49 |
| | | $100.00 | $0.03 | $0.01 | $0.86 | $0.54 |
| | | $105.00 | $0.08 | $0.04 | $0.69 | $0.61 |
| | | $110.00 | $0.03 | -$0.02 | $1.44 | $0.52 |
| | | $115.00 | $0.01 | -$0.16 | $1.18 | $0.36 |
| | 11/17/2023 | $95.00 | $0.00 | -$0.13 | $1.13 | $0.75 |
| | | $97.50 | -$0.02 | -$0.11 | $1.56 | $0.80 |
| | | $100.00 | $0.02 | -$0.11 | $1.04 | $0.80 |
| | | $105.00 | -$0.03 | -$0.01 | $0.88 | $0.72 |
| | | $110.00 | $0.03 | -$0.11 | $1.29 | $0.88 |
| NVDA | 9/15/2023 | $425.00 | $0.73 | $0.55 | $6.61 | $6.15 |
| | | $430.00 | $0.54 | $0.55 | $7.57 | $6.38 |
| | | $435.00 | $0.61 | $0.57 | $7.28 | $6.37 |
| | | $440.00 | $0.68 | $0.62 | $6.79 | $6.37 |
| | | $445.00 | $0.75 | $0.59 | $7.03 | $6.36 |
| | 11/17/2023 | $425.00 | $0.07 | -$0.20 | $8.82 | $7.22 |
| | | $430.00 | $0.18 | -$0.10 | $9.22 | $7.34 |
| | | $435.00 | -$0.26 | -$0.04 | $12.12 | $7.37 |
| | | $440.00 | $0.28 | -$0.04 | $9.11 | $7.27 |
| | | $445.00 | -$0.52 | -$0.12 | $14.60 | $7.52 |
| SPY | 9/15/2023 | $434.00 | $0.10 | $0.03 | $2.10 | $1.55 |
| | | $435.00 | $0.14 | $0.02 | $2.92 | $1.55 |
| | | $436.00 | $0.18 | $0.03 | $2.22 | $1.53 |
| | | $437.00 | $0.03 | $0.03 | $2.84 | $1.51 |
| | | $438.00 | $0.16 | $0.02 | $2.00 | $1.53 |
| | 11/17/2023 | $434.00 | -$0.05 | -$0.47 | $3.09 | $2.45 |
| | | $435.00 | -$0.02 | -$0.45 | $3.21 | $2.44 |
| | | $436.00 | $0.07 | -$0.30 | $3.98 | $2.49 |
| | | $437.00 | -$0.10 | $0.02 | $2.55 | $1.88 |
| | | $438.00 | $0.06 | -$0.46 | $2.83 | $2.47 |

| Exhibit A1: Final Mean P&L Statistics using 10000 Market Calibrated Stochastic Volatility Paths (λ = 3%) | | | | | | |
|---|---|---|---|---|---|---|
| **Symbol** | **Maturity Date** | **Strike** | **RL Mean** | **Delta Mean** | **RL SD** | **Delta SD** |
| AAPL | 9/15/2023 | $165.00 | -$4.31 | -$4.69 | $2.24 | $2.81 |
| | | $170.00 | -$5.84 | -$6.07 | $2.65 | $2.62 |
| | | $175.00 | -$5.66 | -$6.51 | $2.06 | $2.44 |
| | | $180.00 | -$5.12 | -$5.72 | $2.36 | $2.74 |
| | | $185.00 | -$0.40 | -$3.84 | $2.00 | $2.63 |
| | 11/17/2023 | $165.00 | -$7.88 | -$10.75 | $3.22 | $5.00 |
| | | $170.00 | -$10.40 | -$12.04 | $4.28 | $4.79 |
| | | $175.00 | -$10.33 | -$12.52 | $3.64 | $4.79 |
| | | $180.00 | -$11.97 | -$12.52 | $5.32 | $4.84 |
| | | $185.00 | -$6.04 | -$9.34 | $6.60 | $6.00 |
| JNJ | 9/15/2023 | $165.00 | -$3.46 | -$4.57 | $1.78 | $2.84 |
| | | $170.00 | -$4.95 | -$5.81 | $2.65 | $2.59 |
| | | $175.00 | -$6.07 | -$6.53 | $2.24 | $2.41 |
| | | $180.00 | -$4.34 | -$5.70 | $2.64 | $2.75 |
| | | $185.00 | -$1.90 | -$4.44 | $3.27 | $2.97 |
| | 11/17/2023 | $150.00 | -$5.18 | -$6.48 | $3.58 | $4.88 |
| | | $160.00 | -$7.31 | -$8.51 | $4.56 | $5.25 |
| | | $170.00 | -$11.41 | -$11.60 | $5.16 | $4.90 |
| | | $180.00 | -$6.02 | -$8.25 | $6.12 | $6.18 |
| | | $190.00 | -$2.92 | -$4.28 | $4.83 | $5.32 |
| MA | 9/15/2023 | $385.00 | -$12.70 | -$13.85 | $5.46 | $5.77 |
| | | $390.00 | -$12.92 | -$14.52 | $6.08 | $5.51 |
| | | $395.00 | -$11.90 | -$14.65 | $6.12 | $5.51 |
| | | $400.00 | -$10.22 | -$14.30 | $6.56 | $5.68 |
| | | $405.00 | -$4.68 | -$13.05 | $8.14 | $6.04 |
| | 11/17/2023 | $385.00 | -$24.13 | -$25.61 | $10.38 | $11.32 |
| | | $390.00 | -$25.76 | -$27.75 | $9.18 | $10.81 |
| | | $395.00 | -$26.51 | -$28.19 | $8.63 | $10.69 |
| | | $400.00 | -$25.78 | -$23.86 | $14.32 | $12.24 |
| | | $405.00 | -$25.66 | -$28.07 | $12.62 | $10.89 |
| META | 9/15/2023 | $275.00 | -$8.30 | -$9.93 | $4.65 | $4.62 |
| | | $280.00 | -$9.53 | -$10.40 | $4.17 | $4.48 |
| | | $285.00 | -$8.93 | -$10.68 | $4.50 | $4.38 |
| | | $290.00 | -$11.01 | -$10.50 | $4.64 | $4.41 |
| | | $295.00 | -$7.94 | -$9.72 | $7.05 | $4.86 |
| | 11/17/2023 | $275.00 | -$19.91 | -$19.66 | $10.39 | $8.47 |
| | | $280.00 | -$16.83 | -$19.87 | $7.46 | $8.46 |
| | | $285.00 | -$14.81 | -$18.44 | $8.65 | $9.07 |
| | | $290.00 | -$19.66 | -$20.41 | $10.52 | $8.48 |
| | | $295.00 | -$15.05 | -$18.97 | $9.18 | $8.96 |
| MSFT | 9/15/2023 | $305.00 | -$7.28 | -$10.36 | $3.94 | $5.03 |
| | | $310.00 | -$10.60 | -$11.19 | $4.98 | $4.80 |
| | | $315.00 | -$11.47 | -$11.91 | $4.82 | $4.62 |
| | | $320.00 | -$9.97 | -$11.74 | $4.46 | $4.63 |
| | | $325.00 | -$9.37 | -$11.39 | $3.73 | $4.85 |
| | 11/17/2023 | $305.00 | -$18.10 | -$20.60 | $7.38 | $8.94 |
| | | $310.00 | -$20.20 | -$22.31 | $7.70 | $8.90 |
| | | $315.00 | -$20.77 | -$21.84 | $8.58 | $9.10 |
| | | $320.00 | -$19.80 | -$22.30 | $7.87 | $8.89 |
| | | $325.00 | -$19.27 | -$21.78 | $8.29 | $9.18 |
| NKE | 9/15/2023 | $97.50 | -$2.25 | -$2.48 | $1.12 | $1.69 |
| | | $100.00 | -$2.85 | -$3.14 | $1.64 | $1.72 |
| | | $105.00 | -$3.66 | -$3.96 | $1.61 | $1.52 |
| | | $110.00 | -$1.64 | -$2.98 | $2.15 | $1.81 |
| | | $115.00 | $0.01 | -$1.38 | $1.16 | $1.55 |
| | 11/17/2023 | $95.00 | -$3.80 | -$5.04 | $2.48 | $3.11 |
| | | $97.50 | -$4.58 | -$5.83 | $2.52 | $3.15 |
| | | $100.00 | -$5.71 | -$6.42 | $2.33 | $3.05 |
| | | $105.00 | -$5.83 | -$6.25 | $2.98 | $3.31 |
| | | $110.00 | -$6.00 | -$7.28 | $2.67 | $3.02 |
| NVDA | 9/15/2023 | $425.00 | -$15.20 | -$15.66 | $9.24 | $8.72 |
| | | $430.00 | -$17.55 | -$15.92 | $11.28 | $8.80 |
| | | $435.00 | -$13.77 | -$15.94 | $8.35 | $8.68 |
| | | $440.00 | -$16.33 | -$15.96 | $9.94 | $8.79 |
| | | $445.00 | -$15.85 | -$16.10 | $11.24 | $8.87 |
| | 11/17/2023 | $425.00 | -$29.01 | -$30.25 | $13.64 | $13.22 |
| | | $430.00 | -$32.92 | -$30.75 | $15.01 | $13.17 |
| | | $435.00 | -$30.30 | -$30.48 | $16.40 | $13.36 |
| | | $440.00 | -$28.24 | -$30.33 | $15.79 | $13.49 |
| | | $445.00 | -$40.83 | -$30.92 | $21.92 | $13.58 |
| SPY | 9/15/2023 | $434.00 | -$15.06 | -$15.87 | $5.38 | $6.06 |
| | | $435.00 | -$10.92 | -$16.50 | $3.48 | $5.89 |
| | | $436.00 | -$14.54 | -$16.30 | $5.39 | $5.90 |
| | | $437.00 | -$16.27 | -$16.50 | $5.69 | $5.86 |
| | | $438.00 | -$12.39 | -$16.62 | $4.50 | $5.86 |
| | 11/17/2023 | $434.00 | -$26.81 | -$29.63 | $10.04 | $11.76 |
| | | $435.00 | -$31.90 | -$30.09 | $12.16 | $12.15 |
| | | $436.00 | -$27.90 | -$31.47 | $7.96 | $11.50 |
| | | $437.00 | -$25.39 | -$23.92 | $12.83 | $13.30 |
| | | $438.00 | -$27.75 | -$30.49 | $9.83 | $11.83 |

**Exhibit A3: Asset Paths Between 8/17/2023 and 11/17/2023 for Each Symbol**

| Date | AAPL | JNJ | MA | META | MSFT | NKE | NVDA | SPY |
|---|---|---|---|---|---|---|---|---|
| 8/17/2023 | $174.00 | $174.01 | $392.62 | $285.09 | $316.88 | $105.05 | $433.44 | $436.29 |
| 8/18/2023 | $174.49 | $172.49 | $392.17 | $283.25 | $316.48 | $104.81 | $432.99 | $436.50 |
| 8/21/2023 | $175.84 | $167.35 | $393.20 | $289.90 | $321.88 | $102.86 | $469.67 | $439.34 |
| 8/22/2023 | $177.23 | $166.02 | $397.84 | $287.60 | $322.46 | $101.46 | $456.68 | $438.15 |
| 8/23/2023 | $181.12 | $164.53 | $401.06 | $294.24 | $327.00 | $98.75 | $471.16 | $443.03 |
| 8/24/2023 | $176.38 | $165.09 | $397.67 | $286.75 | $319.97 | $97.63 | $471.63 | $436.89 |
| 8/25/2023 | $178.61 | $166.25 | $402.89 | $285.50 | $322.98 | $98.84 | $460.18 | $439.97 |
| 8/28/2023 | $180.19 | $164.29 | $407.44 | $290.26 | $323.70 | $99.63 | $468.35 | $442.76 |
| 8/29/2023 | $184.12 | $164.31 | $411.65 | $297.99 | $328.41 | $101.77 | $487.84 | $449.16 |
| 8/30/2023 | $187.65 | $163.73 | $413.91 | $295.10 | $328.79 | $102.10 | $492.64 | $451.01 |
| 8/31/2023 | $187.87 | $161.68 | $412.64 | $295.89 | $327.76 | $101.71 | $493.55 | $450.35 |
| 9/1/2023 | $189.46 | $160.48 | $415.57 | $296.38 | $328.66 | $102.36 | $485.09 | $451.19 |
| 9/5/2023 | $189.70 | $160.68 | $411.50 | $300.15 | $333.55 | $100.32 | $485.48 | $449.24 |
| 9/6/2023 | $182.91 | $158.01 | $413.18 | $299.17 | $332.88 | $100.18 | $470.61 | $446.22 |
| 9/7/2023 | $177.56 | $160.03 | $414.62 | $298.67 | $329.91 | $97.93 | $462.41 | $444.85 |
| 9/8/2023 | $178.18 | $160.56 | $414.84 | $297.89 | $334.27 | $97.67 | $455.72 | $445.52 |
| 9/11/2023 | $179.36 | $162.66 | $416.69 | $307.56 | $337.94 | $96.79 | $451.78 | $448.45 |
| 9/12/2023 | $176.30 | $163.58 | $416.27 | $301.66 | $331.77 | $96.30 | $448.70 | $445.99 |
| 9/13/2023 | $174.21 | $163.99 | $416.30 | $305.06 | $336.06 | $96.13 | $454.85 | $446.51 |
| 9/14/2023 | $175.74 | $163.74 | $413.34 | $311.72 | $338.70 | $97.19 | $455.81 | $450.36 |
| 9/15/2023 | $175.01 | $161.45 | $414.31 | $300.31 | $330.22 | $96.26 | $439.00 | $443.37 |
| 9/18/2023 | $177.97 | $162.47 | $417.13 | $302.55 | $329.06 | $95.51 | $439.66 | $443.63 |
| 9/19/2023 | $179.07 | $162.20 | $413.53 | $305.07 | $328.65 | $94.62 | $435.20 | $442.71 |
| 9/20/2023 | $175.49 | $162.91 | $410.52 | $299.67 | $320.77 | $94.04 | $422.39 | $438.64 |
| 9/21/2023 | $173.93 | $161.66 | $403.36 | $295.73 | $319.53 | $91.59 | $410.17 | $431.39 |
| 9/22/2023 | $174.79 | $160.50 | $402.22 | $299.08 | $317.01 | $90.85 | $416.10 | $430.42 |
| 9/25/2023 | $176.08 | $160.26 | $402.49 | $300.83 | $317.54 | $90.60 | $422.22 | $432.23 |
| 9/26/2023 | $171.96 | $159.02 | $395.38 | $298.96 | $312.14 | $90.17 | $419.11 | $425.88 |
| 9/27/2023 | $170.43 | $157.11 | $395.48 | $297.74 | $312.79 | $89.42 | $424.68 | $426.05 |
| 9/28/2023 | $170.69 | $156.88 | $399.44 | $303.96 | $313.64 | $89.63 | $430.89 | $428.52 |
| 9/29/2023 | $171.21 | $155.75 | $395.91 | $300.21 | $315.75 | $95.62 | $434.99 | $427.48 |
| 10/2/2023 | $173.75 | $155.15 | $395.85 | $306.82 | $321.80 | $94.56 | $447.82 | $427.31 |
| 10/3/2023 | $172.40 | $155.34 | $391.06 | $300.94 | $313.39 | $95.09 | $435.17 | $421.59 |
| 10/4/2023 | $173.66 | $155.52 | $393.76 | $305.58 | $318.96 | $95.89 | $440.41 | $424.66 |
| 10/5/2023 | $174.91 | $157.14 | $394.20 | $304.79 | $319.36 | $95.79 | $446.88 | $424.50 |
| 10/6/2023 | $177.49 | $157.64 | $397.97 | $315.43 | $327.26 | $97.11 | $457.62 | $429.54 |
| 10/9/2023 | $178.99 | $158.54 | $394.74 | $318.36 | $329.82 | $96.88 | $452.73 | $432.29 |
| 10/10/2023 | $178.39 | $158.36 | $400.37 | $321.84 | $328.39 | $97.62 | $457.98 | $434.54 |
| 10/11/2023 | $179.80 | $156.18 | $399.81 | $327.82 | $332.42 | $98.65 | $468.06 | $436.32 |
| 10/12/2023 | $180.71 | $156.33 | $399.90 | $324.16 | $331.16 | $99.25 | $469.45 | $433.66 |
| 10/13/2023 | $178.85 | $156.85 | $398.03 | $314.69 | $327.73 | $99.91 | $454.61 | $431.50 |
| 10/16/2023 | $178.72 | $157.53 | $401.16 | $321.15 | $332.64 | $102.04 | $460.95 | $436.04 |
| 10/17/2023 | $177.15 | $156.09 | $401.77 | $324.00 | $332.06 | $103.01 | $439.38 | $436.02 |
| 10/18/2023 | $175.84 | $152.73 | $393.21 | $316.97 | $330.11 | $103.77 | $421.96 | $430.21 |
| 10/19/2023 | $175.46 | $152.32 | $387.87 | $312.81 | $331.32 | $103.05 | $421.01 | $426.43 |
| 10/20/2023 | $172.88 | $153.00 | $384.41 | $308.65 | $326.67 | $102.67 | $413.87 | $421.19 |
| 10/23/2023 | $173.00 | $151.39 | $383.67 | $314.01 | $329.32 | $102.81 | $429.75 | $420.46 |
| 10/24/2023 | $173.44 | $151.23 | $386.91 | $312.55 | $330.53 | $105.18 | $436.63 | $423.63 |
| 10/25/2023 | $171.10 | $151.57 | $386.31 | $299.53 | $340.67 | $103.54 | $417.79 | $417.55 |
| 10/26/2023 | $166.89 | $149.00 | $364.59 | $288.35 | $327.89 | $100.02 | $403.26 | $412.55 |
| 10/27/2023 | $168.22 | $145.60 | $364.08 | $296.73 | $329.81 | $97.98 | $405.00 | $410.68 |
| 10/30/2023 | $170.29 | $147.03 | $372.42 | $302.66 | $337.31 | $101.80 | $411.61 | $415.59 |
| 10/31/2023 | $170.77 | $148.34 | $376.35 | $301.27 | $338.11 | $102.77 | $407.80 | $418.20 |
| 11/1/2023 | $173.97 | $148.69 | $377.82 | $311.85 | $346.07 | $100.88 | $423.25 | $422.66 |
| 11/2/2023 | $177.57 | $150.24 | $382.69 | $310.87 | $348.32 | $105.08 | $435.06 | $430.76 |
| 11/3/2023 | $176.65 | $151.34 | $386.05 | $314.60 | $352.80 | $107.06 | $450.05 | $434.69 |
| 11/6/2023 | $179.23 | $151.70 | $386.16 | $315.80 | $356.53 | $107.25 | $457.51 | $435.69 |
| 11/7/2023 | $181.82 | $150.90 | $388.87 | $318.82 | $360.53 | $109.36 | $459.55 | $436.93 |
| 11/8/2023 | $182.89 | $150.35 | $389.70 | $319.78 | $363.20 | $109.39 | $465.74 | $437.25 |
| 11/9/2023 | $182.41 | $147.42 | $387.96 | $320.55 | $360.69 | $107.00 | $469.50 | $433.84 |
| 11/10/2023 | $186.40 | $147.25 | $394.38 | $328.77 | $369.67 | $106.11 | $483.35 | $440.61 |
| 11/13/2023 | $184.80 | $147.63 | $394.35 | $329.19 | $366.68 | $104.20 | $486.20 | $440.19 |
| 11/14/2023 | $187.44 | $147.66 | $397.65 | $336.31 | $370.27 | $105.75 | $496.56 | $448.73 |
| 11/15/2023 | $188.01 | $148.80 | $396.83 | $332.71 | $369.67 | $107.82 | $488.88 | $449.68 |
| 11/16/2023 | $189.71 | $150.10 | $397.10 | $334.19 | $376.17 | $107.61 | $494.80 | $450.23 |
| 11/17/2023 | $189.69 | $149.79 | $400.30 | $335.04 | $369.85 | $105.96 | $492.98 | $450.79 |

| Exhibit A4: Final P&L's using Empirical Asset Paths | | | | |
|---|---|---|---|---|
| **Symbol** | **Maturity Date** | **Strike** | **RL Mean** | **Delta Mean** |
| AAPL | 9/15/2023 | $165.00 | -$2.49 | -$2.08 |
| | | $170.00 | -$4.44 | -$3.97 |
| | | $175.00 | -$6.54 | -$8.19 |
| | | $180.00 | -$11.78 | -$12.66 |
| | | $185.00 | $0.05 | -$12.95 |
| | 11/17/2023 | $165.00 | -$6.02 | -$8.76 |
| | | $170.00 | -$7.62 | -$12.97 |
| | | $175.00 | -$11.69 | -$13.04 |
| | | $180.00 | -$16.86 | -$15.72 |
| | | $185.00 | -$6.00 | -$10.20 |
| JNJ | 9/15/2023 | $165.00 | -$1.55 | -$3.66 |
| | | $170.00 | -$6.20 | -$3.40 |
| | | $175.00 | -$3.48 | -$3.28 |
| | | $180.00 | -$1.92 | -$2.03 |
| | | $185.00 | -$0.22 | -$1.34 |
| | 11/17/2023 | $150.00 | $1.17 | -$2.06 |
| | | $160.00 | -$6.78 | -$6.41 |
| | | $170.00 | -$4.14 | -$5.06 |
| | | $180.00 | -$0.89 | -$1.42 |
| | | $190.00 | -$0.96 | -$1.72 |
| MA | 9/15/2023 | $434.00 | -$2.93 | -$4.65 |
| | | $435.00 | -$2.07 | -$5.48 |
| | | $436.00 | -$7.64 | -$6.92 |
| | | $437.00 | -$12.02 | -$8.35 |
| | | $438.00 | -$10.22 | -$9.48 |
| | 11/17/2023 | $434.00 | -$18.85 | -$21.05 |
| | | $435.00 | -$13.08 | -$19.98 |
| | | $436.00 | -$15.16 | -$19.85 |
| | | $437.00 | -$15.90 | -$14.78 |
| | | $438.00 | -$17.74 | -$17.09 |
| META | 9/15/2023 | $97.50 | -$2.33 | -$3.51 |
| | | $100.00 | -$6.22 | -$5.08 |
| | | $105.00 | -$7.59 | -$6.23 |
| | | $110.00 | -$6.81 | -$7.28 |
| | | $115.00 | -$8.44 | -$8.57 |
| | 11/17/2023 | $95.00 | -$12.21 | -$9.64 |
| | | $97.50 | -$15.33 | -$11.04 |
| | | $100.00 | -$14.36 | -$12.37 |
| | | $105.00 | -$16.96 | -$14.25 |
| | | $110.00 | -$16.72 | -$12.31 |
| MSFT | 9/15/2023 | $305.00 | -$3.67 | -$3.70 |
| | | $310.00 | -$3.92 | -$3.75 |
| | | $315.00 | -$5.41 | -$5.84 |
| | | $320.00 | -$7.67 | -$7.61 |
| | | $325.00 | -$3.64 | -$10.50 |
| | 11/17/2023 | $305.00 | -$10.50 | -$11.62 |
| | | $310.00 | -$14.42 | -$14.31 |
| | | $315.00 | -$16.11 | -$16.65 |
| | | $320.00 | -$17.75 | -$17.60 |
| | | $325.00 | -$19.02 | -$17.36 |
| NKE | 9/15/2023 | $385.00 | -$0.59 | -$2.39 |
| | | $390.00 | -$3.04 | -$3.48 |
| | | $395.00 | -$1.99 | -$2.90 |
| | | $400.00 | -$0.41 | -$0.78 |
| | | $405.00 | $0.11 | -$0.31 |
| | 11/17/2023 | $385.00 | -$5.57 | -$6.33 |
| | | $390.00 | -$6.09 | -$9.25 |
| | | $395.00 | -$10.09 | -$10.84 |
| | | $400.00 | -$2.44 | -$2.16 |
| | | $405.00 | -$3.34 | -$2.81 |
| NVDA | 9/15/2023 | $425.00 | $5.14 | $3.53 |
| | | $430.00 | $1.48 | $5.58 |
| | | $435.00 | $9.39 | $6.43 |
| | | $440.00 | $12.26 | $6.53 |
| | | $445.00 | $2.82 | $4.45 |
| | 11/17/2023 | $425.00 | -$14.64 | -$17.57 |
| | | $430.00 | -$18.30 | -$18.28 |
| | | $435.00 | -$17.45 | -$17.35 |
| | | $440.00 | -$11.40 | -$20.09 |
| | | $445.00 | -$18.57 | -$12.90 |
| SPY | 9/15/2023 | $275.00 | -$9.92 | -$8.94 |
| | | $280.00 | -$7.08 | -$10.35 |
| | | $285.00 | -$11.21 | -$10.62 |
| | | $290.00 | -$9.82 | -$11.75 |
| | | $295.00 | -$11.86 | -$12.88 |
| | 11/17/2023 | $275.00 | -$27.37 | -$36.30 |
| | | $280.00 | -$35.85 | -$38.74 |
| | | $285.00 | -$23.99 | -$29.21 |
| | | $290.00 | -$26.78 | -$23.93 |
| | | $295.00 | -$32.30 | -$45.58 |